\documentclass[12pt,preprint]{aastex}
\usepackage{graphicx}
\usepackage{pdflscape}
\usepackage{color}

\shorttitle{Photometry and spectroscopy of multiple populations along the AGB of NGC\,2808 and NGC\,6121 (M\,4)} 
\shortauthors{A.\,F.\,Marino, et al.\ } 
\usepackage{ulem}

\newcommand \kmsec {\mbox{km~s$^{\rm -1}$}}
\newcommand \teff {T$_{\rm eff}$}
\newcommand \logg {\mbox{log~{\it g}}}
\newcommand \vmicro {\mbox{$\xi_{\rm t}$}}

\begin{document}
\title{Spectroscopy and photometry of multiple populations along the asymptotic-giant branch of NGC\,2808 and NGC\,6121 (M\,4) \footnote{           Based on observations with  the                               NASA/ESA {\it Hubble Space Telescope},
                               obtained at  the Space Telescope Science
                               Institute,  which is operated by AURA, Inc.,
                               under NASA contract NAS 5-26555.}
}

\author{
A.\ F.\,Marino\altaffilmark{2}, 
A.\ P.\,Milone\altaffilmark{2},
D.\ Yong\altaffilmark{2},
G.\ Da Costa\altaffilmark{2},
%J.\ Anderson\altaffilmark{3},
M.\ Asplund\altaffilmark{2},
L.\ R.\,Bedin\altaffilmark{3},
H.\ Jerjen\altaffilmark{2},
D.\ Nardiello\altaffilmark{4},
G.\ Piotto\altaffilmark{3,4},
A.\ Renzini\altaffilmark{3},
M.\ Shetrone\altaffilmark{5} %,
%P.\ B.\,Stetson\altaffilmark{6}
}

\altaffiltext{2}{Research School of Astronomy \& Astrophysics, Australian National University, Canberra, ACT 2611, Australia}
\altaffiltext{3}{Istituto Nazionale di Astrofisica - Osservatorio Astronomico di Padova, Vicolo dell'Osservatorio 5, Padova, IT-35122}
\altaffiltext{4}{Dipartimento di Fisica e Astronomia 'Galileo Galilei' - Univ. di Padova, Vicolo dell'Osservatorio 3, Padova, IT-35122}
\altaffiltext{5}{McDonald Observatory, The University of Texas at Austin, 1 University Station, C1400, Austin, TX 78712-0259, USA}
 
\begin{abstract}
  We present a photometric and spectroscopic study of multiple
  populations along the asymptotic-giant branch (AGB) of the
   intermediate-metallicity globular clusters (GCs) NGC\,2808 and NGC\,6121
  (M\,4). 
  Chemical abundances of O, Na, Mg, Al, Si, Ca, Sc, Ti, V, Cr, Fe, Co,
  Ni, Zn, Y, and Ce in AGB stars from high-resolution
  FLAMES$+$UVES@VLT spectra are reported for both clusters.
  Our spectroscopic results have been combined with multi-wavelength
  photometry from the {\it HST} UV survey of Galactic GCs and
  ground-based photometry, plus proper motions derived 
  by combining stellar positions from ground-based images and Gaia DR1.
  Our analysis reveals that the AGBs of both clusters host
  multiple populations with different chemical composition. 

  In M\,4 we have identified two main populations of stars with different
  Na/O content, lying on distinct 
  AGBs in the $m_{\rm F438W}$ vs.\,$C_{\rm F275W, F336W, F438W}$ and
  the $V$ vs.\,$C_{\rm U,B,I}$ pseudo-CMDs.

  In the more massive and complex GC NGC\,2808 
  three groups of stars with different chemical abundances 
occupy
  different locations on the so-called "chromosome map'' photometric
  diagram. 
The spectroscopic+photometric comparison of stellar populations along the AGB and
  the red giants of this GC suggests that the AGB hosts stellar
  populations with a range in helium abundances spanning from
  primordial up to high contents of $Y\sim 0.32$. 
On the other hand, from our dataset, there is no evidence for stars with extreme helium
  abundance ($Y \sim  0.38$) on the AGB, suggesting that the most
  He-rich stars of NGC\,2808 do not reach this phase. 
 \end{abstract}


\section{Introduction}
\label{introduction}
It is now established that virtually all Galactic Globular
Clusters (GCs) host two or more stellar populations with different
chemical composition: a first population of stars with the same
chemical abundance as halo field stars at similar metallicity, and
second population(s) of stars enhanced in helium, nitrogen, and sodium
and depleted in carbon and oxygen (e.g.\, Kraft\,1994; Gratton et al.\,2012).  
In the last decade, multiple stellar populations have been homogeneously
studied in a large number of Galactic GCs by using both
spectroscopy and photometry (e.g.\,Carretta et al.\,2009; Piotto et
al.\,2015; Milone et al.\,2017 and references therein).
 Most of these studies focused on stars along the red-giant
branch (RGB), the sub-giant branch, and the main sequence (MS).
Little attention, however, has been devoted to more evolved stages,
like the asymptotic-giant branch (AGB). 

Spectroscopic work has shown that AGB stars of the few studied GCs can
exhibit star-to-star variation in light elements, in analogy
with what is observed along the RGB. These conclusions are based both
on early analysis of molecular bands (e.g.\,Smith \& Norris\,1993)
and, more recently,
on high-resolution spectroscopy of light elements (e.g.\,Ivans et
al.\,1999,\,2001; Johnson et al.\,2015). 

On the photometric side, appropriate combination of ultraviolet and
optical filters of the {\it Hubble Space Telescope} ({\it HST\,}),
such as the $C_{\rm F275W,F336W,F438W}=$($m_{\rm F275W}-m_{\rm
  F336W}$)$-$($m_{\rm F336W}-m_{\rm F438W}$) pseudo-color, have been
proved to provide a powerful means to identify multiple populations
along the entire color-magnitude diagram (CMD) of GCs, from the MS up
to the RGB and the horizontal-branch (Milone et al.\,2013).   
 Recent work has then used this index
% $m_{\rm F438W}$ vs.\,$C_{\rm F275W,F336W,F438W}$ 
to investigate the AGB, and found that the AGB of NGC\,2808 hosts
 three distinct sequences (Milone et al.\,2015a). Similarly, the
 AGBs of the GCs NGC\,7089 and NGC\,6352 are found to be inconsistent with a simple
 stellar population (Milone et al.\,2015b; Nardiello et al.\,2015a;
 for further photometric evidence see Gruyters et al.\,2017).  

 On the other hand, there is some evidence that the population
  ratio, in terms of light elements chemical abundances, is different
  among RGB and AGB stars. 
Early analysis on cyanogen (CN) band
  strengths by Smith \& Norris\,(1993) found a different distribution
  of CN-band strengths among AGBs and RGBs in 
  NGC\,6752 and M\,5. 
More recently, a lack of AGB stars with the highest sodium abundance
observed on the RGB has been shown for NGC\,6752 and
  NGC\,6266 (M\,62, Campbell et al.\,2013; Lapenna et al.\,2015,
  2016). In 47~Tuc, the AGB displays similar Na abundance variations as
  in the RGB, but a fraction of $\lesssim$20\% of Na-rich RGB stars
  may not reach the AGB phase (Johnson et al.\,2015). These results
  suggest that some stars enriched in the high temperature H-burning
  products fail to ascend the AGB
  and have raised new interest in stellar populations in this
  evolutionary phase.

As the chemical enrichment in Na is indicative for a star to
  belong to a second stellar population in GCs, this means that the AGBs of some GCs
  may display a paucity of second population stars compared to the RGB.  
 Qualitatively, enrichment in helium among second population
 stars, as observed in GCs (e.g.\,Milone et al.\,2014), can account for the
 lack of the Na(and He)-richest stars along the AGB. Indeed, He-enhanced
 stars have smaller envelope masses (and higher surface temperature)
 on the horizontal branch, and, if the mass is low enough, they evolve
 directly to the white dwarf sequence through the so-called AGB-manque' phase. 
The lack of AGB stars in M\,62 is consistent with this scenario
 as the populous second-generation of this cluster 
  has extremely high helium abundance (Y$\sim$0.33, Milone\,2015). 
  Quantitatively, most GCs have internal helium variations, $\Delta (Y)$,
   of just a few hundredths (Milone\,2015).
Evolution models of low-mass stars suggest that, besides helium, the maximum sodium content
expected on the AGB is a function of both metallicity and age,
as with younger ages stars are more massive on the horizontal branch
hence will have a better chance to climb the AGB even if helium rich
(e.g., Charbonnel \& Chanterau\,2016).

In this paper we combine multi-wavelength photometry from {\it HST}
and ground-based telescopes with high-resolution spectroscopy from the
Ultraviolet and Visual Echelle Spectrograph (UVES, Dekker et al.\,2000) of the Very Large
Telescope (VLT) to further investigate the multiple populations along the
AGB of the midly metal-poor GCs NGC\,2808 and NGC\,6121 (M\,4). 

The targets of this paper, NGC\,2808 and M\,4, have been widely
investigated in the context of multiple populations. NGC\,2808 is one
of the most massive and complex GCs of the Milky Way. Its ``chromosome
map'', a photometric diagram introduced by Milone et al.\,(2015a) to
separate different stellar groups in GCs, hosts at
least five distinct populations
(e.g.\,Milone et al.\,2015a; Carretta\,2015). 
It exhibits stellar populations with very-high helium
content ($Y \sim 0.32$ and $Y \sim 0.38$, D'Antona et al.\,2005;
Piotto et al.\,2007; Marino et al.\,2014) and extreme variations in light element
abundances, including O, Mg, Al, Si, and Na (Carretta et al.\,2009; Carretta\,2014).  
 
In contrast, M\,4 is a much simpler GC. It hosts ``only'' two main populations  (Marino et
al.\,2008; Lee et al.\,2009; Milone et al.\,2014, 2017) with
relatively small variations in light elements (Ivans et al.\,1999;
Marino et al.\,2008, 2011; Carretta et al.\,2009; Villanova et
al.\,2011) and helium ($\Delta Y \sim 0.02$, Villanova et al.\,2012;
Nardiello et al.\,2015b) compared to NGC\,2808. 

 The AGBs of both NGC\,2808 and M\,4 have been recently investigated by
 means of high-resolution spectroscopy. Wang et al.\,(2016) has
 determined the sodium abundance of 31 AGB stars in NGC\,2808 and concluded that
 this cluster hosts second-generation AGB stars and that the fraction
 of Na-rich AGB stars is {\it higher} than that observed on the RGB. 

The presence of multiple populations along the AGB of M\,4 is a quite
controversial issue.  The chemical composition of 15 AGB stars in M\,4
has been recently 
investigated by  MacLean et al.\,(2016) using the 2dF$+$HERMES
facility on the Anglo-Australian Telescope. These authors have
suggested that the AGB is mostly populated by stars with low sodium
and high oxygen abundance.
However, these findings have been photometrically challenged by the recent
 work of Lardo et al.\,(2017) who have analyzed the $V$ vs.\,$C_{\rm
   U,B,I}=$($U-I$)$-$($B-I$) diagram of M\,4 and concluded that its
 broadened AGB is not consistent with a simple population. 

The layout of the paper is as follows:
Sect.~\ref{data} presents the photometric and spectroscopic data that
we have analysed; our chemical analysis is described in
Sect.~\ref{sec:chemical_analysis}; our results on chemical abundances
of AGB stars are discussed in Sect.~\ref{chimica} in conjunction with the
photometric properties; finally Sect.~\ref{summary} is a summary of
our results.

\section{Observations and data reduction}\label{data}
In order to study multiple stellar populations along the AGB of M\,4 and
NGC\,2808 we have combined information from both photometry and spectroscopy.
 The photometric and spectroscopic datasets will be described in the following subsections. 

\subsection{The photometric dataset and target selection}\label{sec:phot}
We have used both {\it HST} and ground-based photometry of NGC\,2808
and M\,4. The photometric and astrometric catalogs from {\it HST} have
been published by Piotto et al.\,(2015) and include stars in the
innermost ($\sim 2.7\arcmin \times 2.7\arcmin$) clusters
regions. They have been derived from images collected through the
$F275W$, $F336W$, $F438W$ filters of the Ultraviolet and Visual Channel of the Wide
Field Camera 3 (UVIS/WFC3) on {\it HST}. The data include stellar proper
motions. Moreover, they also provide photometry of images from 
Anderson et al.\ (2008), collected with
the F814W band of the Wide Field Channel of the Advanced Camera for
Survey on {\it HST}. Only stars that,
according to their proper motions, are considered cluster members are
included in the analysis. 
 This photometric dataset has been used in several works (Piotto et
 al.\,2015; Milone et al.\,2015a, 2017; Simioni et al.\,2016) to
 investigate stellar populations in NGC\,2808 and M\,4 and we refer the reader 
 to these papers for further details on the data and the data reduction.  

In addition, we have used the wide-field photometric catalogs from the
database maintained by Peter Stetson which are derived from images collected
with ground-based facilities (see Stetson 1987, 1994, 2000; Stetson et
al.\, 2014). These data have been previously used by Monelli et
al.\,(2013) to study multiple stellar populations along the RGB of
both M\,4 and NGC\,2808. We refer the reader to these papers for
details on the dataset and on the method used to  
derive the photometry and astrometry. 
We have calculated relative proper motions by combining the positions
of stars derived from images collected with the Wide Field Imager
(WFI) of the MPI 2.2m telescope in La Silla with those from the Gaia
data release 1 (Lindegren et al.\,2016). 

The WFI data used to calculate stellar proper motions in the field of
M\,4 consist of 6$\times$100s$+$3$\times$10s images 
collected on June 18$^{\rm th}$ and 21$^{\rm th}$, 2001 through the
$B$ band (program 69.D-05282). Photometry and astrometry of these
images have been computed using the programs from Anderson et
al.\,(2006) while proper 
motions have been derived as in Anderson et al.\,(2006) and Piotto et
al.\,(2012). 
Both {\it HST} and ground-based photometry has been corrected for
differential reddening following the recipe by Milone et al.\,(2012). 

\begin{figure}[ht!]
   \centering
   \epsscale{0.604}
   \plotone{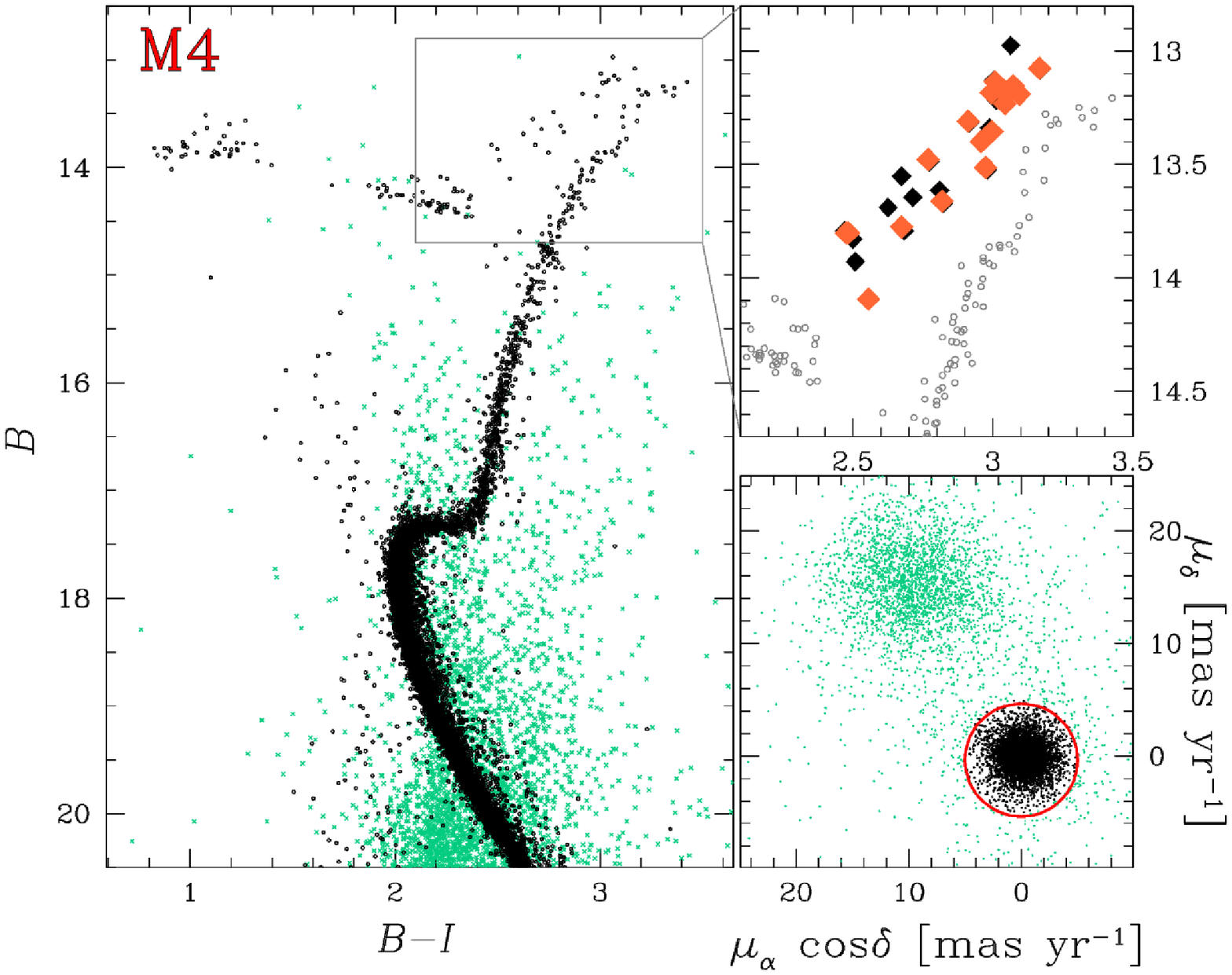}
   \epsscale{0.368}
   \plotone{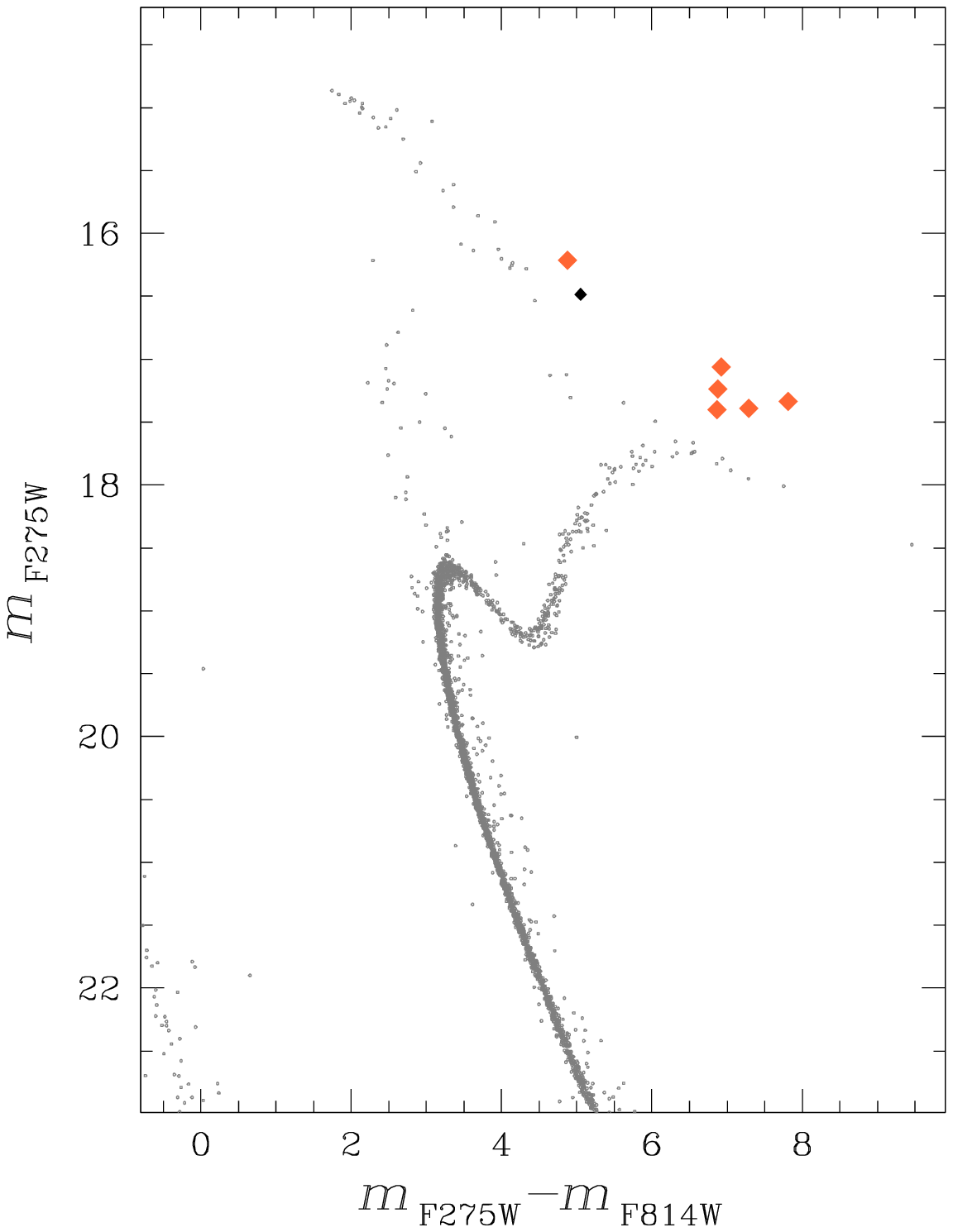}
   \caption{\textit{Left panels:} $B$ vs.\,$B-I$ CMD of stars with
     radial distance less than 20 arcmin from the center of M\,4
     (Stetson\, et al.\,2014). Cluster members and field stars are
     colored black and aqua, respectively, and have been selected on
     the basis of their proper motions. The vector-point diagram of
     stellar proper motions obtained by combining information from Gaia
     catalogues and WFI images is plotted on the lower-right side of
     the CMD. We also show a zoom of the CMD of cluster members around
     the AGB and the RGB.  
     \textit{Right panel:} $m_{\rm F275W}$ vs.\,$m_{\rm F275W}-m_{\rm
       F814W}$ CMD of NGC\,2808 cluster members from {\it HST}
     photometry (Piotto et al.\,2015). These AGB stars in the central
     {\it HST} field are also included in the ground-based
     photometry. The photometrically-selected AGB stars are indicated
     with filled diamonds in both CMDs, and those observed also
     spectroscopically are colored orange.   
   }
   \label{fig:M4target}
   \end{figure}

\begin{figure}[ht!]
   \centering
   \epsscale{0.604}
   \plotone{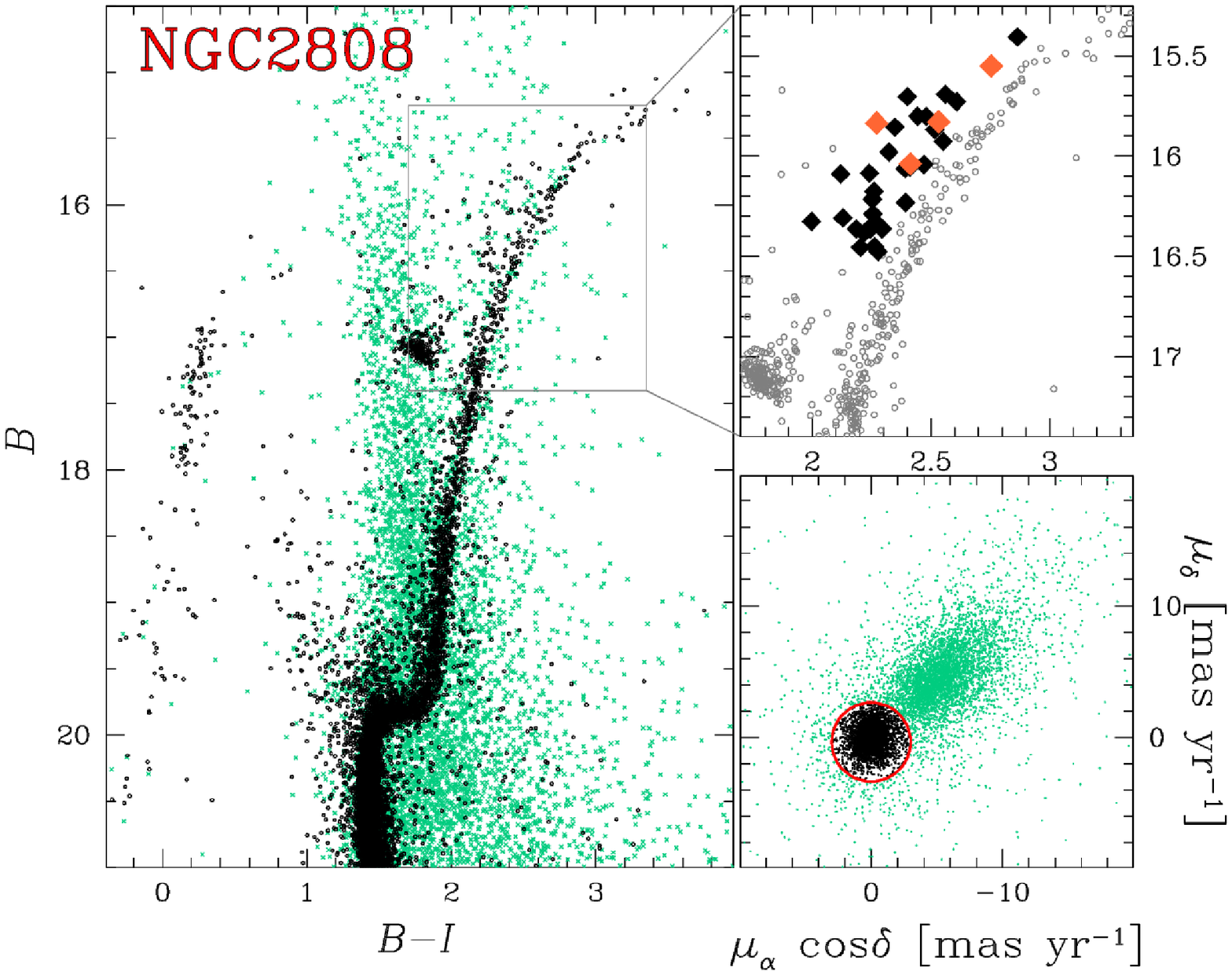}
   \epsscale{0.368}
   \plotone{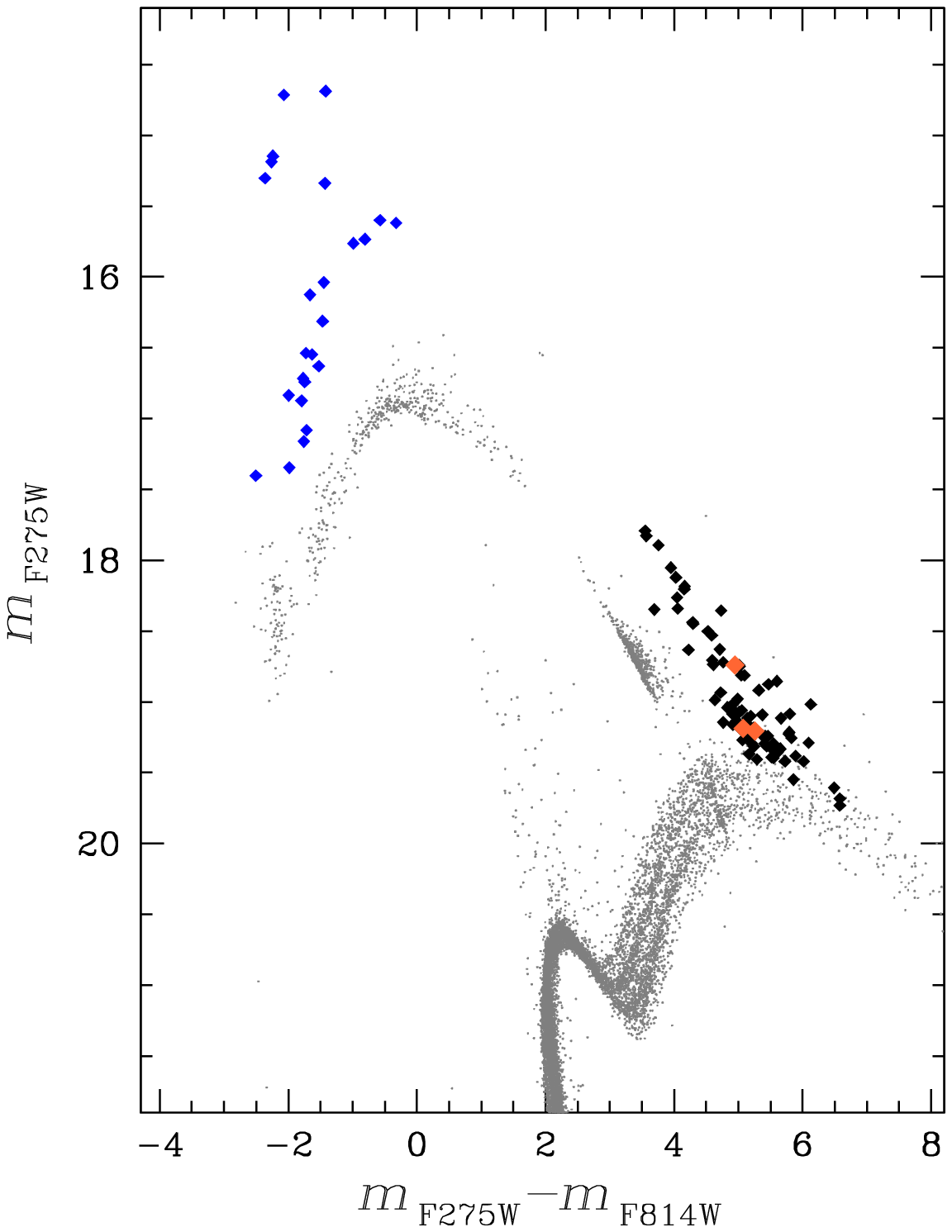}
   \caption{As in Figure~\ref{fig:M4target} but for NGC\,2808. In this
     case the CMD from ground-based photometry includes only stars
     with radial distance between 1.5 and 12.0 arcmin from the cluster
     center, and does not include the AGB observed in the central {\it
     HST} field. The blue points plotted in the right-panel CMD of
   cluster members mark candidate AGB-manqu{\'e} stars.}
   \label{fig:NGC2808target}
   \end{figure}

The CMDs of stars in the field of view of  M\,4 and NGC\,2808 from
ground-based and {\it HST} photometry are shown in
Fig.~\ref{fig:M4target} and Fig.~\ref{fig:NGC2808target}, where we
indicate with orange diamonds our sample of seventeen and seven
spectroscopically-analysed AGB
stars, respectively. The remaining AGB stars have been studied from
photometry only, and are represented with black diamonds. We also show
the vector point diagram of proper motions obtained by combining the
coordinates of stars from the Gaia and WFI catalogs, that has been
used to identify candidate cluster members and field stars.  
Our target AGB stars were carefully selected from CMDs derived from
both the ground-based and {\it HST} observations. Specifically, for NGC\,2808 we have selected
three stars on the three main AGB sequences identified by Milone et
al.\,(2015a) by using the $C_{\rm
  F275W,F336W,F438W}$ index from $HST$ photometry.  
The remaining four stars of NGC\,2808 have been selected from ground-based photometry by Stetson (2000). 
In the case of M\,4, multi-wavelength {\it HST} photometry from Piotto
et al.\,(2015) is available for six AGB stars, while the remaining
eleven stars have been identified only on the CMD obtained from
ground-based photometry. 

\subsection{The spectroscopic dataset}

Our dataset consists of FLAMES-UVES spectra (RED580 setting, Pasquini
et al.\,2000) collected under the
programs 093.D-0789 and 094.D-0455. 
Final spectra have been obtained by co-adding 2$\times$2775s
  and 30$\times$2775s exposures for M\,4 and NGC\,2808, respectively.  
Data were reduced using the UVES pipelines (Ballester et al.\,2000),
including bias subtraction, flat-field correction, wavelength
calibration, sky subtraction and spectral rectification. 
The spectra have a spectral coverage of $\sim$2000~\AA\, with the
central wavelength at $\sim$5800~\AA. 
Telluric subtraction has been performed by using the ESO MOLECFIT tool (Smette et al.\,2015; Kaush et al.\,2015).
The typical signal to noise ratio for the final combined spectra
around the [O\,{\sc i}]$\lambda$6300~\AA\ line ranges from S/N$\sim$130 to
$\sim$230 for M\,4, and from $\sim$160 to $\sim$250 for NGC\,2808.

Radial velocities (RVs) were derived using
the iraf@FXCOR task, which cross-correlates the object spectrum with a
template. For the template we used a synthetic spectrum obtained
through MOOG (Sneden 1973), computed with a model stellar atmosphere interpolated from the
Castelli \& Kurucz (2004) grid, adopting parameters
(effective temperature/surface gravity/microturbulence/metallicity) =
(4900~K/2.0/2.0~\kmsec\/$−$1.20). 
Each spectrum was corrected to the rest-frame system, and observed RVs were then
corrected to the heliocentric system. 
The mean heliocentric RVs of M\,4 and NGC\,2808 are 
$\langle$RV$\rangle$=$+69.5\pm0.8$~\kmsec\ ($\sigma$=3.2~\kmsec) and
$\langle$RV$\rangle$=$+96.7\pm3.9$~\kmsec\ ($\sigma$=9.6~\kmsec), respectively.
These values agree with those obtained from RGB stars in the same clusters, e.g.\,
$\langle$RV$\rangle$=$+$70.6$\pm$1~\kmsec\ for M\,4 (Marino et al.\,2008), and
$\langle$RV$\rangle$=$+$102.4$\pm$1~\kmsec\ for NGC\,2808 ($\sigma$=9.8~\kmsec, Carretta et al.\,2006).
As discussed in Sect.~\ref{sec:phot}, our spectroscopic targets had
already passed the membership selection criterion based on proper motions. In
the end, 
all proper-motion members are also RV members, with the exception of
M4 star 16235035-2632478 (Table~1).
Based on the chemical abundances, discussed in the next sections, the membership is further
confirmed by the fact that all the target stars have [Fe/H] consistent
with the cluster mean metallicity, except 1623035-2632478 which was not analyzed.

%__________________________________________________________________
%__________________________________________________________________

%__________________________________________________________________
%__________________________________________________________________

\section{Chemical abundance analysis}
\label{sec:chemical_analysis}
Chemical abundances have been derived from a local thermodynamical
equilibrium (LTE) analysis using MOOG (version 2013; Sneden 1973), plus
$\alpha$-enhanced model atmospheres of Castelli \& Kurucz
(2004). The line list and reference solar abundances are as in Marino et
al.\,(2008). 

To infer the atmospheric parameters used in our chemical analysis, we
took advantage of the high-resolution and high-S/N of our UVES spectra
and employed Fe lines. 
Specifically: (i) effective temperatures (\teff) were derived by
imposing the excitation potential equilibrium of the Fe\,{\sc i}
lines; (ii) surface gravities (\logg) were set with the ionisation equilibrium
between Fe\,{\sc i} and Fe\,{\sc ii} lines, but allowing Fe\,{\sc ii}
abundances to be slightly higher than Fe\,{\sc
  i} to take into account deviations from LTE (Lind et al.\,2012;
Bergemann et al.\,2012); (iii) microturbulent velocities (\vmicro) were
set to minimize any dependence on Fe\,{\sc i} abundances on
equivalent widths (EWs). 

In Table~1 we list our adopted spectroscopic parameters, together
with \teff\ obtained from the Alonso $(B-V)$-\teff\ calibrations (Alonso et
al.\,1999), assuming a reddening of E$(B-V)$=0.36 (Harris 1996) and
E$(B-V)$=0.19 (Bedin et al.\,2000) for M\,4 and NGC\,2808,
respectively, and \logg\ derived from the canonical relation assumed
mass equal to 0.60~$\rm {M_{\odot}}$, $(m-M)_V$=12.82 for M\,4,
and $(m-M)_V$=15.59 for NGC\,2808, temperatures, and
apparent $V$ magnitudes as above. 
By comparing the spectroscopically-derived atmospheric parameters with those
from photometry we get mean differences of
$<$\teff$_{B-V}-$\teff$_{\rm {Fe}}>$=$-12\pm15$~K, rms=72~K, and
$<$\logg$_{phot}-$\logg$_{\rm {Fe}}>=-0.09\pm$0.04, rms=0.21.
The fact that our adopted \teff\ and \logg\ average values agree reasonably
with the photometric ones based on recent
reddening values gives us confidence in our adopted \teff\ scales.
On the other hand, we notice that for M\,4 AGBs, the discrepancy with the
photometric temperatures is larger for hotter stars; specifically for
stars with \teff$>$4800~K, the photometric temperatures and gravities are lower by
$>$100~K and $>$0.2, respectively\footnote{The star with the highest discrepancy (star
  16234268-2631209) has also a spectrum with broader lines, compared to
  the other stars, and, for this reason, we suggest caution with interpretation of this star.}.
A variation in temperature of $\pm$100~K will change photometric
surface gravities by around $\pm$0.05-0.06~dex only. Hence the discrepancy in
\teff\ alone is not enough to account for the lower photometric gravity in the
higher-temperature stars. By changing the assumed mass by
$\pm$0.1~$\rm {M/M_\odot}$, \logg\ varies by around $\pm0.08$~dex.
We are not able to explain why hotter stars display larger discrepancy
between spectroscopic and photometric parameters in M\,4, which is not
observed in NGC\,2808, but note that in the latter case the observed range in
\teff\ is smaller ($\Delta$\teff=460~K vs.\,$\Delta$\teff=790~K in M\,4).

To test the magnitude of non-LTE effects on our metallicity
  values, assumed to be equal to Fe~{\sc i} abundances, 
we derived the non-LTE corrections to the Fe~{\sc i} spectral lines from Lind et al.\,(2012)
by using the inspect
tool\footnote{http://inspect.coolstars19.com/index.php?n=Main.HomePage}
for one RGB (\#20766) and one AGB (\#16233142-2633110) star in M\,4. 
We found that both the RGB and the AGB stars
have positive non-LTE correction, 0.06 and 0.08~dex respectively.
Hence, metallicities in non-LTE should be $\sim$0.08~dex higher in our
AGB sample.
Note that the AGB non-LTE correction is only
marginally higher (by $0.021\pm0.002$) than that of the RGB.

For all the elements, but O and Al, chemical abundances
were obtained from the equivalent widths derived from Gaussian fitting
of isolated spectral lines. Oxygen and aluminum were derived from
spectral synthesis of the lines [O\,{\sc i}] $\lambda$6300~\AA\ and
$\lambda$6363~\AA\ and Al doublet $\lambda$6697~\AA\ to account for
the blending with other spectral features.
In the end, we have been able to infer chemical abundances for sixteen
elements, namely O, Na, Mg, Al, Si, Ca, Sc (Sc\,{\sc i} and Sc\,{\sc
  ii}), Ti (Ti\,{\sc i} and Ti\,{\sc ii}), V, Cr  (Cr\,{\sc i} and
Cr\,{\sc ii}), Fe  (Fe\,{\sc i} and Fe\,{\sc ii}), Co, Ni, Zn, Y\,{\sc
ii} and Ce\,{\sc ii}.

Estimates of the uncertainties in chemical abundances have been
obtained by re-running the abundances varying
\teff/\logg/[m/H]/\vmicro, one at a time, by
$\pm$100~K/$\pm$0.20/$\pm$0.15/$\pm$0.30~\\\kmsec.
The uncertainties used in \teff\ and \logg\ are reasonable as
suggested by the comparison with the photometric values discussed
above. As internal errors in [m/H] and \vmicro\ we conservatively
adopt $\pm$0.15~dex and $\pm$0.30~\kmsec. 
In addition to the contribution introduced by internal errors in
atmospheric parameters, we estimated the contribution ($\sigma_{\rm fit}$) due to the finite S/N, which
affects the measurements of EWs and the spectral synthesis. The
contribution due to EWs has been calculated by varying the EWs of
spectral lines by $\pm$4.5m\AA. This value has been derived
by comparing EWs from various exposures of the same stars. 
The variations in the abundances obtained by varying the EWs have been
then divided by $\sqrt{(N-1)}$ (where $N$ is the number of available
spectral lines).
For the elements analysed through spectral synthesis we estimated the
error in their chemical abundances by varying the continuum placement
in the synthesis within a reasonable range. 
Variations in chemical abundances due to each contribution,
plus the total error estimate obtained by summing in quadrature all
the different contributions are listed in Tab.~\ref{tab:err}.

The average abundances and the corresponding uncertainties, defined as
the rms from different spectral lines for the same species, are listed in 
Tables~\ref{tab:abu1}--\ref{tab:abu3}.

\section{The chemical composition of multiple populations along the AGB}
\label{chimica}

An illustration of our derived chemical abundances for AGB stars in
M\,4 and NGC\,2808 is plotted in Fig.~\ref{fig:box:m4,n2808}.
The most obvious difference we observe among the AGBs of the two GCs
is a much higher range in Al for NGC\,2808. M\,4 AGBs exhibit 
higher mean abundances for O, Mg, Si, Zn, and the neutron-capture
elements Y and Ce.
In analogy to what is observed among RGB stars, significant scatter
is observed in both clusters for the elements involved in the
high-temperature H-burning, namely O, Na; Al and Mg in the case of NGC\,2808.
This fact demonstrates that the AGBs of both M\,4 and NGC\,2808 host
multiple stellar populations.
Direct evidence of AGB stars with different Na and O abundances in M\,4 is
further provided in Fig.~\ref{fig:spettri}, where we compare the
spectra of two stars with similar atmospheric parameters but
different [O/Fe] and [Na/Fe].
We note that the [Fe/H] values are all consistent with cluster
membership. The higher spreads in the Zn, Y\,{\sc ii} and Ce\,{\sc ii} abundances
are most likely just the result of the small number of lines and
larger errors. Indeed, the error estimates for these elements listed in
Table~\ref{tab:err} are relatively high, though we cannot exclude that they are
over-estimated as they are higher than our observed rms values.

%__________________________________________________________________
   \begin{figure}[ht!]
   \centering
   \epsscale{0.99}
   \plotone{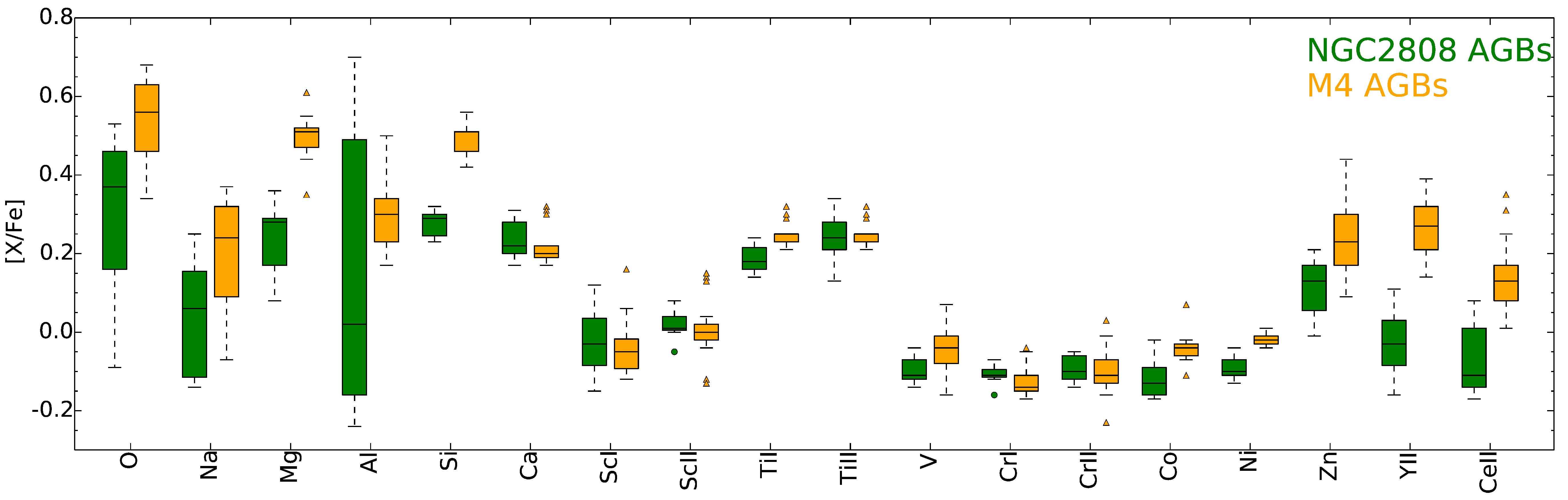}
   \caption{Box-and-whisker plot for the elemental abundances of AGB
     stars in NGC\,2808 (green) and M\,4 (orange). 
     Each box represents the inner quartile of the distribution, and
     the median abundance is marked by the horizontal line. The
     whiskers include the 99.3\% of the data, outliers are plotted by triangles.}
   \label{fig:box:m4,n2808}
   \end{figure}
%__________________________________________________________________

A visual comparison between AGB and RGB abundances in the two analysed
clusters is shown in Fig.~\ref{fig:box}. This comparison reveals that
AGB and RGB exhibit similar distribution for most elements.   
The comparison does not extend beyond Ni because Zn, Y~{\sc ii} and Ce~{\sc ii}
abundances are not available for the RGB stars studied in Marino et
al.\,(2008) and Carretta\,(2014, 2015).

%__________________________________________________________________
   \begin{figure}[ht!]
   \centering
   \epsscale{0.8}
   \plotone{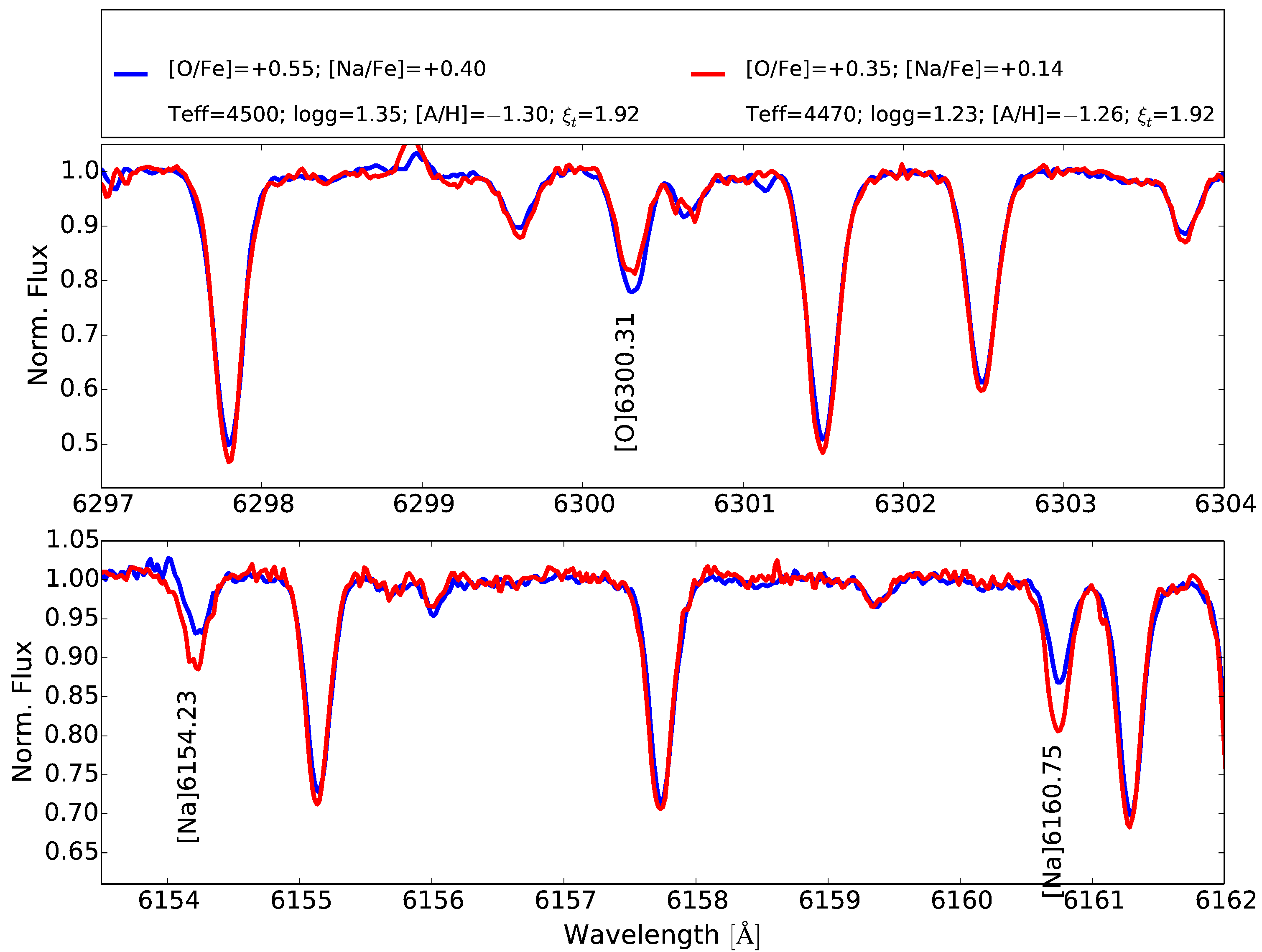}
   \caption{Comparison between the spectra of the AGB stars 16234740-2631463
     (blue) and 16233667-2630397 (red), in M\,4. These stars have similar atmospheric
     parameters but different sodium and oxygen abundance.} 
   \label{fig:spettri}
   \end{figure}
%__________________________________________________________________

Remarkable exceptions are the distributions of oxygen, sodium,
magnesium and aluminum for NGC\,2808, for which we observe differences in range between RGB and AGB stars.
To properly compare these variations, we have calculated the difference, $\delta$, between the
90$^{\rm th}$ and the 10$^{\rm th}$ percentile of the distribution of
these elements for both RGB and AGB stars. We have also associated to
each measurement an uncertainty that has been calculated by means of
bootstrapping as in Milone et al.\,(2014) and which is indicative of the
robustness of the $\delta$ determinations. For NGC\,2808 a
large and significant difference between the $\delta$ values for RGB and AGB stars
has been derived for oxygen, ($\delta_{\rm RGB}^{\rm
  [O/Fe]}=0.94\pm0.03$, $\delta_{\rm AGB}^{\rm [O/Fe]}=0.60\pm0.09$.  
For Na, Mg and Al the differences in $\delta$ values are also large, but the significance is lower
($\delta_{\rm RGB}^{\rm [Na/Fe]}=0.58\pm0.05$, $\delta_{\rm
  AGB}^{\rm [Na/Fe]}=0.40\pm0.16$),  ($\delta_{\rm RGB}^{\rm
  [Mg/Fe]}=0.46\pm0.18$, $\delta_{\rm AGB}^{\rm
  [Mg/Fe]}=0.18\pm0.05$), and ($\delta_{\rm RGB}^{\rm
  [Al/Fe]}=1.13\pm0.13$, $\delta_{\rm AGB}^{\rm
  [Al/Fe]}=0.77\pm0.40$).  
Further, as is apparent from Fig.~\ref{fig:box}, the
results indicate that in NGC\,2808 stars with the largest
abundance of Na and Al and the lowest O and Mg content are clearly
absent in the
analyzed sample of AGB stars. This suggests that the stellar population
with extreme chemical composition in  NGC\,2808 avoids the AGB
phase. 

In contrast, the distributions of sodium and oxygen for RGB and AGB
stars in M\,4 are quite similar, even though the mean Na for AGBs is a
bit lower \footnote{Note that Marino et al.\,(2008) applied an average non-LTE
  correction of $-$0.02~dex from Gratton et al.\,(1999) to the Na abundances of RGB
  stars, which are smaller than more recent computations by Lind et
  al.\,(2011). Hence, to not introduce systematics in our RGB-AGB
  comparison we do not apply any non-LTE correction to our AGB
  abundances. We warn the reader that, by comparing the RGB Na
  abundances from Marino et al.\,(2008) with those derived here, an
  additional small difference by 0.02~dex could be present, making the
mean Na relative to Fe of RGB stars $+$0.29, instead of
$+$0.27~dex.}. 
For this GC, we find $\delta_{\rm RGB}^{\rm
  [O/Fe]}=0.26\pm0.02$ and $\delta_{\rm AGB}^{\rm [O/Fe]}=0.24\pm0.06$
for oxygen, and $\delta_{\rm RGB}^{\rm [Na/Fe]}=0.43\pm0.02$,
$\delta_{\rm AGB}^{\rm [Na/Fe]}=0.37\pm0.07$ for sodium. 
Further, there is no indication for any abundance distribution
differences between the RGB and AGB stars for the other elements
measured; this is also the case for NGC~2808 except for the
differences already mentioned. 

In the next sub-sections we present results on the light-element
(anti-)correlations, and on the connection between
chemical abundances and photometric properties of AGBs in M\,4
(Sect.~\ref{sub:AGBm4}) and NGC\,2808 (Sect.~\ref{sub:AGBn2808}) in
the context of multiple stellar populations.

%__________________________________________________________________
   \begin{figure}[ht!]
   \centering
   \epsscale{0.8}
   \plotone{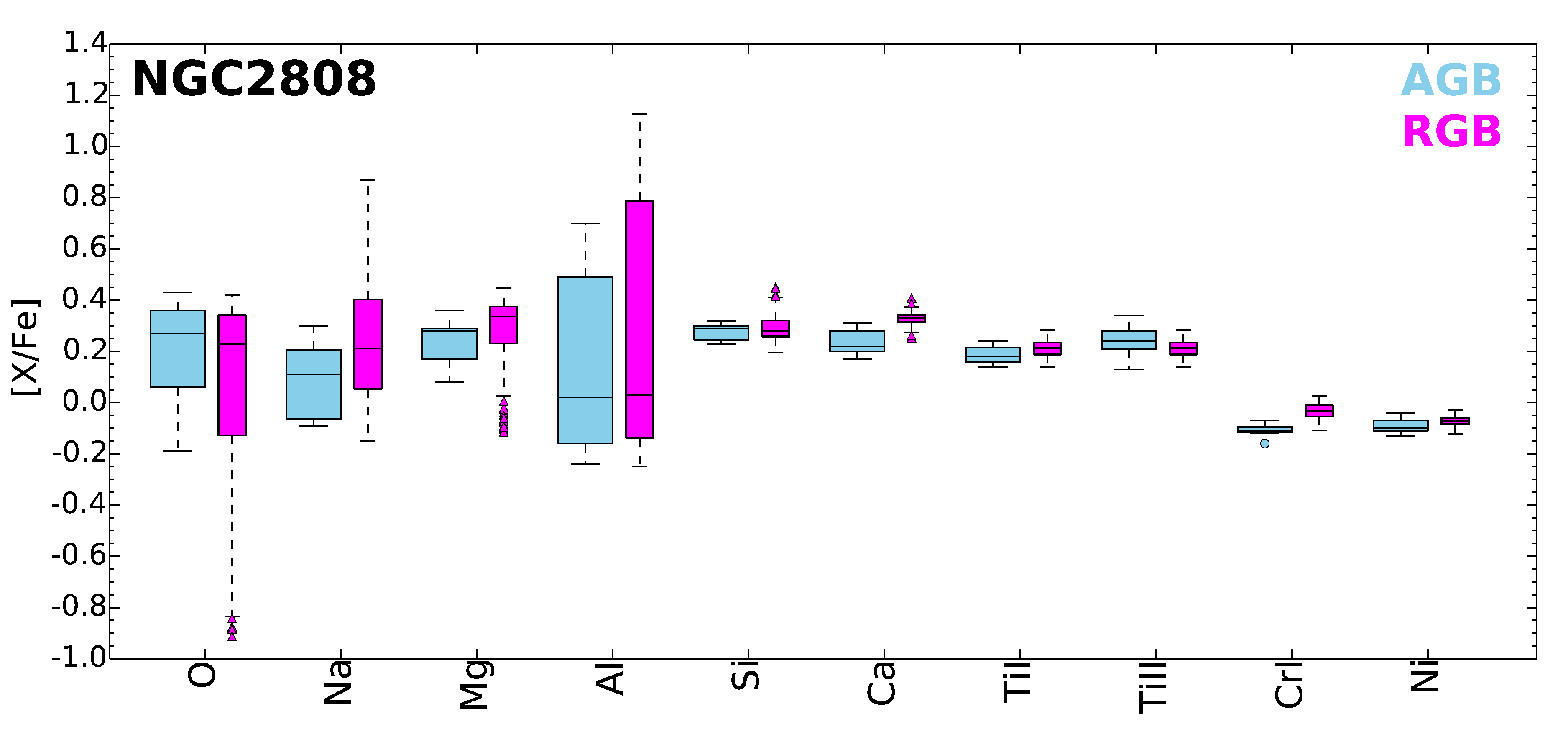}
   \plotone{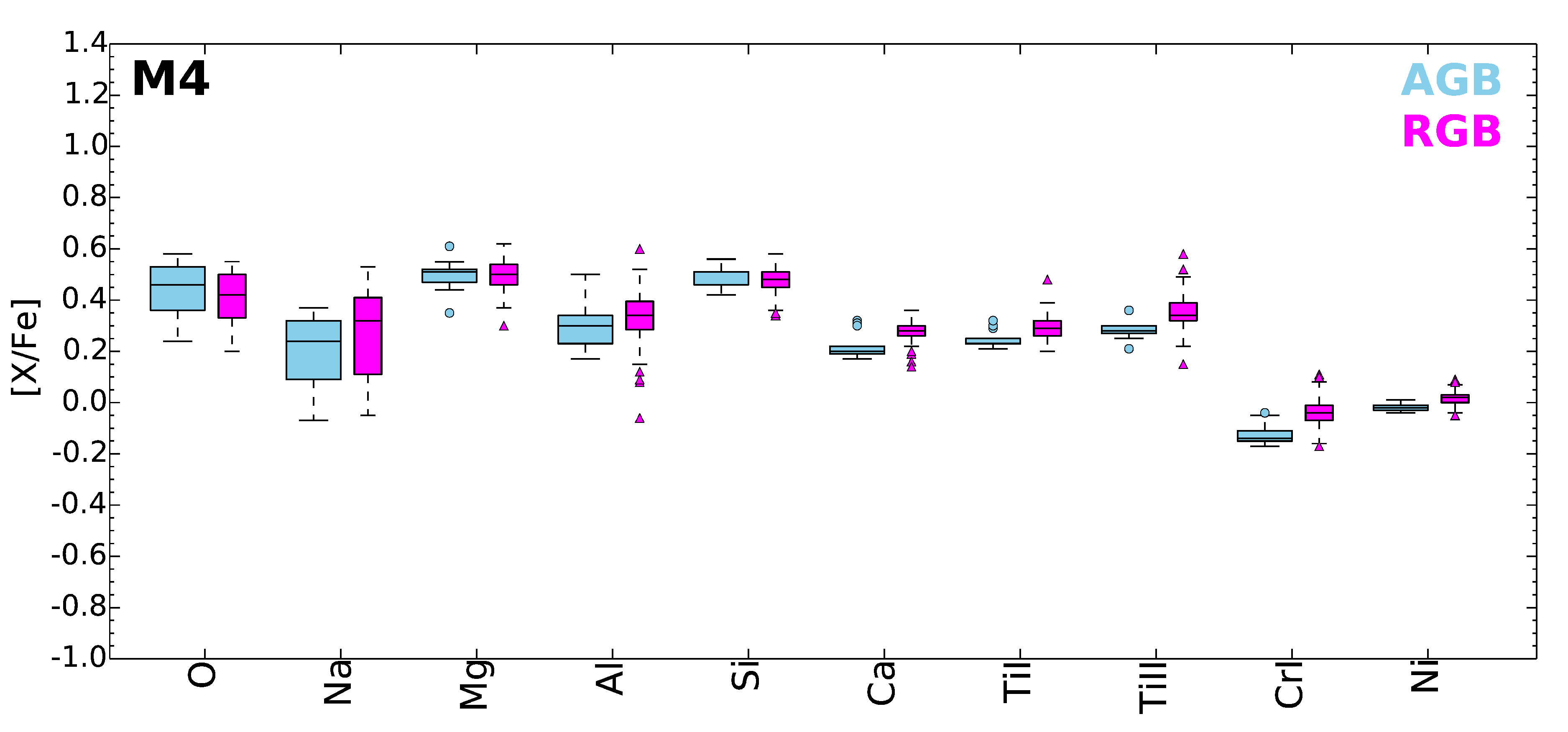}
   \caption{Box-and-whisker plot for the elemental abundance of AGB
     and RGB stars in NGC\,2808 (upper panel) and M\,4 (lower panel),
     using the same vertical scale to emphasize the difference among
     the two clusters.
     The meaning of the boxes, whiskers and outliers is as in Fig.~\ref{fig:box:m4,n2808}.
      Abundances for RGB stars of NGC\,2808 and M\,4 are from Carretta
      (2014, 2015) and Marino et al.\,(2008), respectively.
    Due to the higher Al solar abundance used in this work, [Al/Fe]
    abundances for RGB stars both in M\,4 and NGC\,2808 have been
    shifted by $-$0.21~dex.} 
   \label{fig:box}
   \end{figure}
%__________________________________________________________________

\subsection{The AGB of M\,4}
\label{sub:AGBm4}
\begin{figure}[ht!]
   \centering
   \epsscale{.33}
   \plotone{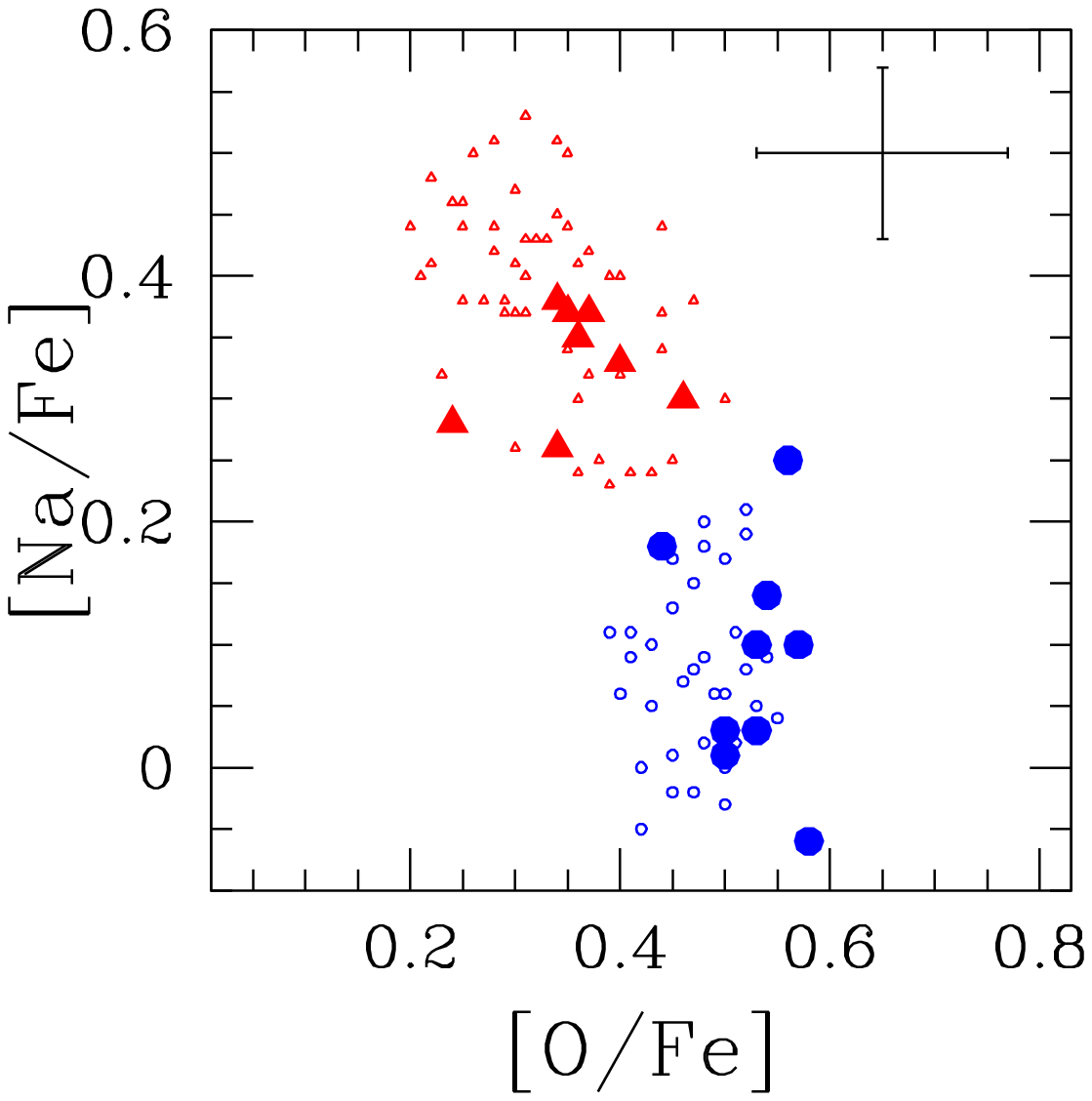}
      \plotone{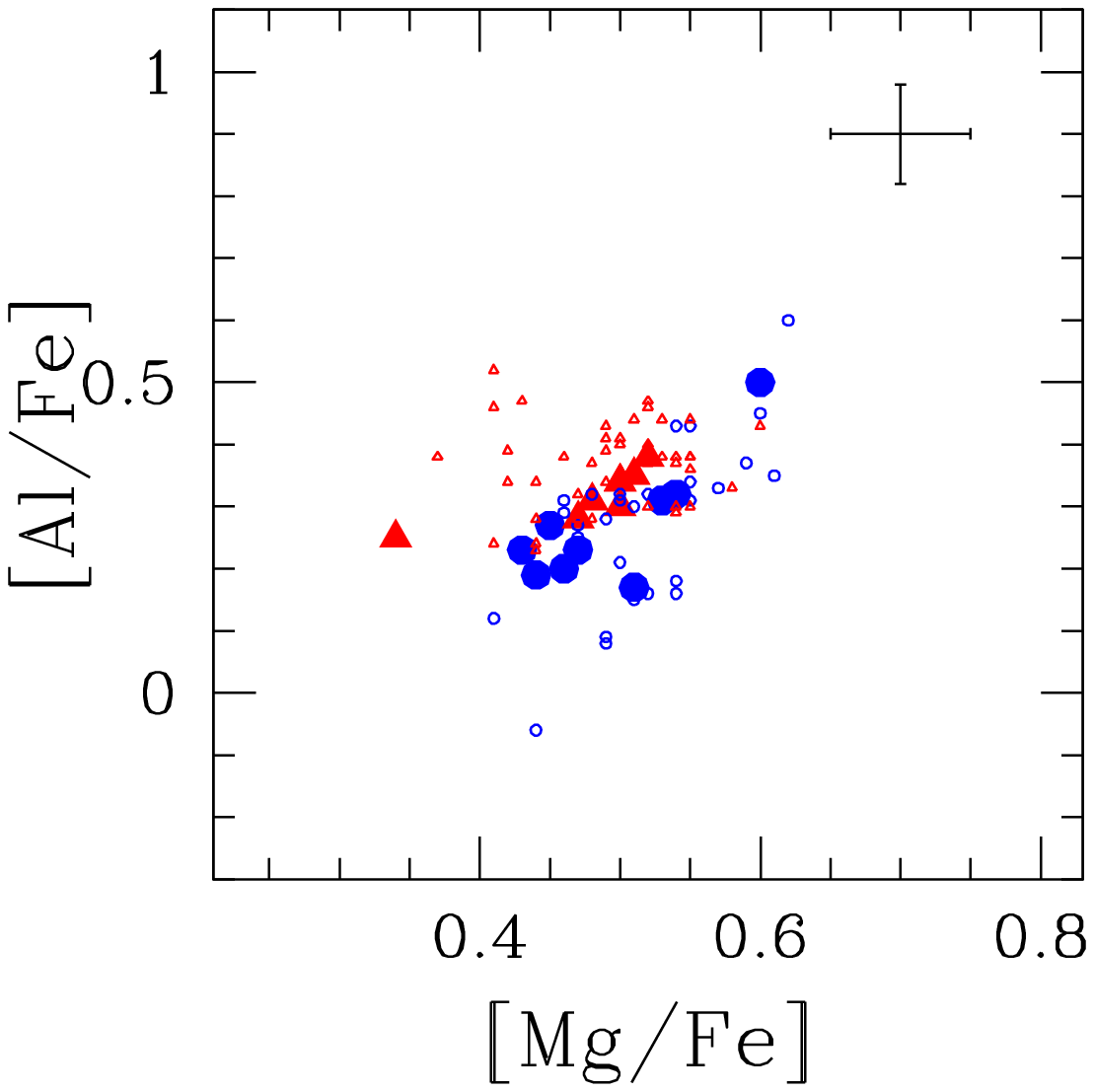}
      \plotone{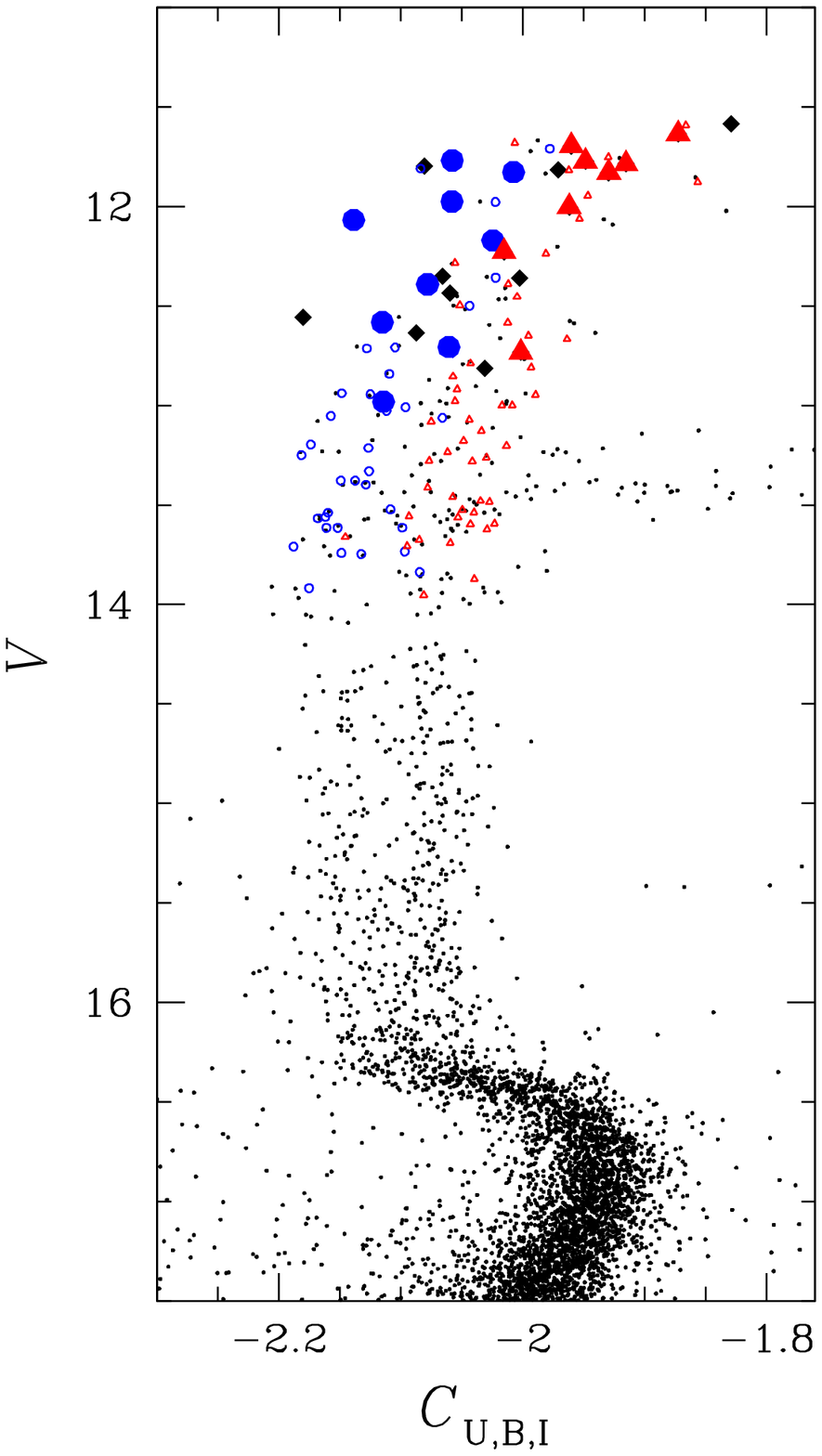}
         \plotone{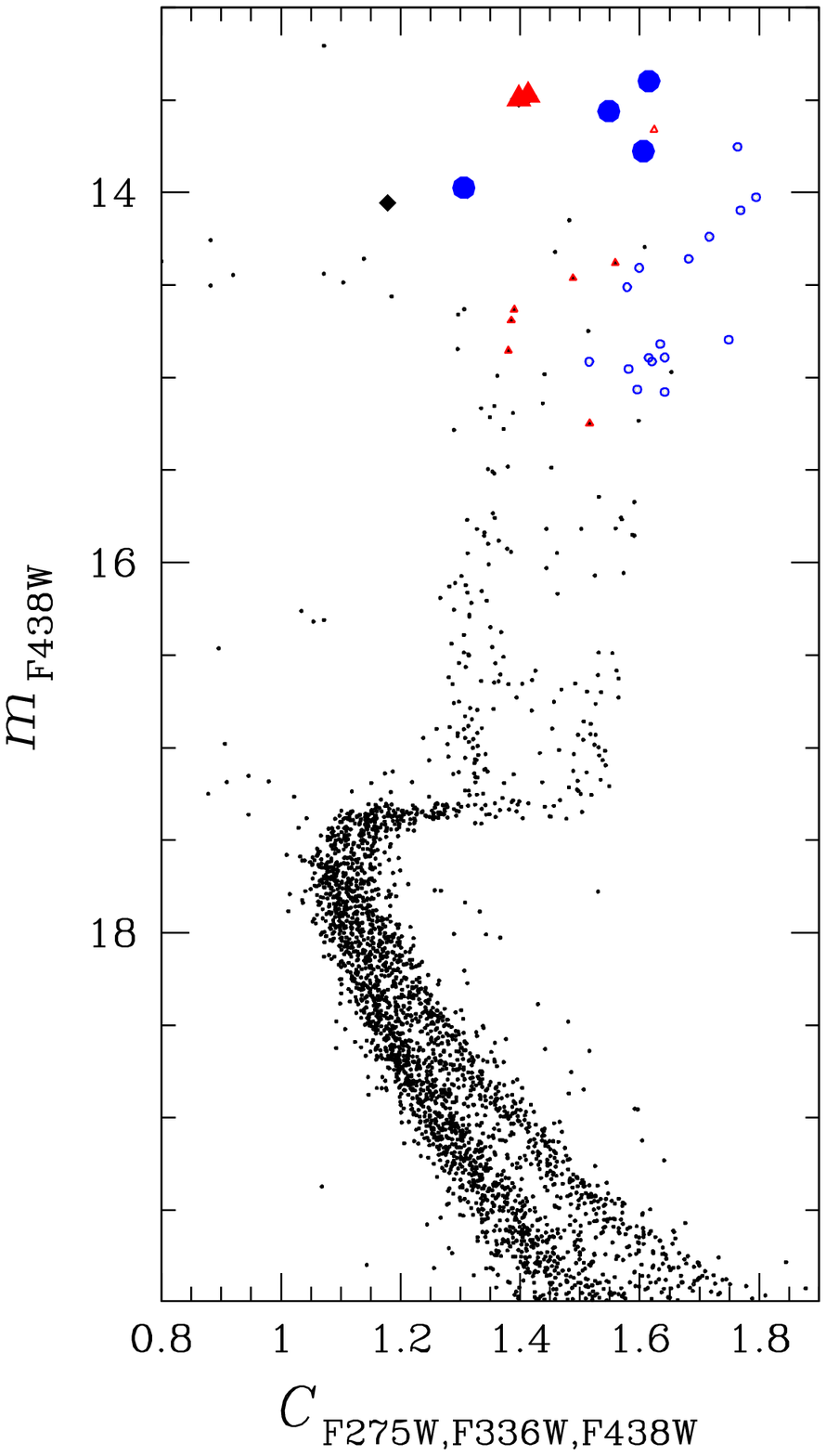}
         \caption{ \textit{Upper panels:} Sodium vs.\,oxygen (left),
           aluminum vs.\,magnesium (right) for AGB stars in M\,4. The
           lower panels show the $V$ vs.\,$C_{\rm U,B,I}$ (left) and
           $m_{\rm F438W}$ vs.\,$C_{\rm F275W,F336W,F438W}$ (right)
           pseudo-CMDs for M\,4 stars.  
           AGB and RGB stars analyzed spectroscopically in this paper
           and in Marino et al.\,(2008) are represented with large and
           small symbols, respectively. The two populations of
           Na-poor/O-rich and Na-rich/O-poor RGB and AGB stars are shown as
           blue filled circles and red filled triangles, respectively, 
           while the remaining AGB
           stars that have been not analyzed spectroscopically are
           represented with black diamonds in the lower-panel
           diagrams. To put Al abundances on the same scale for AGBs
           and RGBs, we subtract 0.21 to the RGB abundances (see Fig.~\ref{fig:box}).
             } 
   \label{fig:lightM4}
   \end{figure}

The upper panels of Figure~\ref{fig:lightM4} compare the position in
the [Na/Fe] vs.\,[O/Fe] and the [Mg/Fe] vs.\,[Al/Fe] plane  of the AGB
  stars analyzed in this paper and the RGB stars from Marino et
  al.\,(2008). AGB stars of M\,4 clearly exhibit the Na-O
  anticorrelation, similar with what has been observed along the RGB,
  and both the RGB and the AGB clearly show two groups of
  Na-poor/O-rich and Na-rich/O-poor stars, although the AGB stars could
  either not reach the highest Na abundances observed on the RGB or
  could be shifted to lower values.  
 There is no evidence for a Mg-Al anticorrelation neither along the
 RGB, nor the AGB, but we note that Na-rich AGB and RGB are slightly
 more-Al rich than Na-poor stars. 
The fractions of Na-poor/O-rich RGB and AGB stars are the same within
one sigma. Specifically, the 42$\pm$5\% of RGB stars (37 out 88
stars) and the 53$\pm$13\% of AGB (9 out 17 stars) are considered
Na-poor. 

 In contrast to our results, MacLean et al.\,(2016) suggested that
  in their sample of M\,4 AGB stars only ‘first generation’ stars were
  present, and that as a result, the M\,4 AGB lacks ‘second
  generation’ O-depleted, Na-rich stars.  There are 10 stars in common
  between our M4 AGB-star sample and that of MacLean et
  al.\,(2016). For these stars we find that our Na abundances are
  0.09$\pm$0.01~dex (standard error of the mean) higher and our O
  abundances 0.06$\pm$0.03 dex lower than the abundances listed in
  MacLean et al.\,(2016).  These differences are consistent with the
  slightly different stellar parameters adopted: our temperatures are
  65~K cooler ($\sigma$ 55~K) and gravities 0.13~dex lower ($\sigma$
  0.19~dex) than their values; and with different analysed spectral
  features: they use the 777~nm O triplet and the 568~nm Na doublet,
  while we use the forbidden O lines, and both the Na doublets at
  $\sim$568~nm and $\sim$616~nm. Intriguingly, if we plot the [Na/Fe]
  and [O/Fe] values from MacLean et al.\,(2016) against each other
  solely for these 10 AGB stars, they clearly fall into two distinct
  groups, as do our abundances for the same stars (see
  Fig.~\ref{fig:lightM4}, upper left panel), with one group having
  lower [O/Fe] and higher [Na/Fe] than the other. The same stars
  occupy each group in both samples, with one group have mean ([O/Fe],
  [Na/Fe]) of (0.55, $-$0.05) and the second having (0.45, 0.22),
  respectively, with the MacLean et al.\,(2016) abundances.  This
  suggests the O-Na anti-correlation is indeed present in the MacLean
  et al. AGB star sample. The mean abundance differences
  between the two groups are essentially identical to that with our
  abundances: ($\Delta$[O/Fe], $\Delta$[Na/Fe]) is ($-$0.10, $+$0.27) for the
  MacLean et al. abundances and ($-$0.17, $+$0.26) for those
  presented here.  We suggest that the conclusion of MacLean et
  al.\,(2016) that the M4 AGB lacks second generation stars may not be
  valid.

 The apparent discrepancy with MacLean et al. might be reconciled by
possible systematically lower Na in the AGB, than in the RGB
counterpart. Our sample seems to suggest a systematic in the same
direction, though much less pronounced. A similar phenomenon was
already noticed by Ivans et al.\,(1999) for Na, and by Smith \&
Norris\,(1993) for the indices S(3839) (mostly sensitive to N) and
W(G) (sensitive to C). Both studies report internal abundance
dispersion on AGB, as on the RGB, but different abundances. Smith \&
Norris discussed possible causes for these discrepancies, including the
C$\rightarrow$N processing occurring in the RGB envelopes which could
make CN-strong red giants evolving to ``less CN-enriched''
AGBs. Furthermore, the same authors showed, through synthetic spectra
computations, that the CN-band strengths on the AGB are
less-pronounced than in the RGB due to different atmospheric
parameters, causing the lower S3839 indices observed in AGB stars. This
discussion probably enlightens that Na and O abundances might be better
indicators of the primordial AGB abundances, although there could be some
effects also on these elements. We suggest that star-to-star elemental
internal variations are a much more reliable tool when we attempt a comparison
between AGB and RGB stars in the context of multiple stellar populations. 

To further compare the AGB with the RGB multiple populations pattern
in M\,4, we take advantage of our photometry. 
It is well known that the two groups of RGB stars with different
 chemical compositions populate distinct sequences in the CMD or
 pseudo-CMD of M\,4 made with an appropriate combination of
 ultraviolet and optical filters (e.g.\,Marino et al.\,2008; Monelli
 et al.\,2013). 
 In the lower-right panel of Figure~\ref{fig:lightM4} we reproduce the
 $m_{\rm F438W}$ vs.\,$C_{\rm F275W,F336W,F438W}$ pseudo-CMD of M\,4
 from Piotto et al.\,(2015), where the two main populations of M\,4
 are clearly visible along the MS and the RGB. 
 The two groups of Na-rich and Na-poor stars selected in the
  upper-left panel clearly correspond to the two main RGBs observed in
 this pseudo-CMD that we have obtained from {\it HST} photometry. 
 We note that AGB stars span a much wider interval in $C_{\rm
   F275W,F336W,F438W}$ than what we expect from observational errors
 only. Moreover, although {\it HST} photometry is available only for
 six stars spectroscopically-analyzed in this paper, Na-rich AGB stars have smaller
 $C_{\rm F275W,F336W,F438W}$ values than Na-poor AGB stars with the
 same luminosity in close analogy with what has been observed along
 the RGB. 

 The lower-left panel of Figure~\ref{fig:lightM4} plots the $V$ vs.\,$C_{\rm U,B,I}$
 %\cubi\ 
  pseudo-CMD from ground-based photometry.
  This diagram has been recently used by Lardo et al.\,(2017) to show
  that both first- and second-generation stars climb the AGB of M\,4. 
Our diagram confirms that the $C_{\rm U,B,I}$
%\cubi\ 
spread for AGB is much
larger than what we expect from photometric uncertainties only and is
 comparable to the $C_{\rm U,B,I}$
 %\cubi\ 
 spread of RGB stars with similar
 luminosity. 
  As already shown by Monelli et al.\,(2013), the two populations of
  Na-rich and Na-poor RGB stars distribute along the two distinct RGB $C_{\rm U,B,I}$
  %\cubi\
 sequences of this cluster. The populations of Na-poor/O-rich and
 Na-rich/O-poor AGB stars exhibit a similar behaviour occupying
 different a location  on the $V$ vs.\,$C_{\rm U,B,I}$
 %\cubi\ 
 plane. 
  These facts confirm that both the $C_{\rm F275W,F336W,F438W}$ and the $C_{\rm U,B,I}$
%\cubi\ 
pseudo-colors are efficient to identify multiple
 stellar populations along the AGB of GCs and suggest that stars
 of both main populations of M\,4 climb the AGB. 
  
 %__________________________________________________________________
%__________________________________________________________________

\subsection{The AGB of NGC\,2808}
\label{sub:AGBn2808}
\begin{figure}[ht!]
   \centering
   \epsscale{.405}
   \plotone{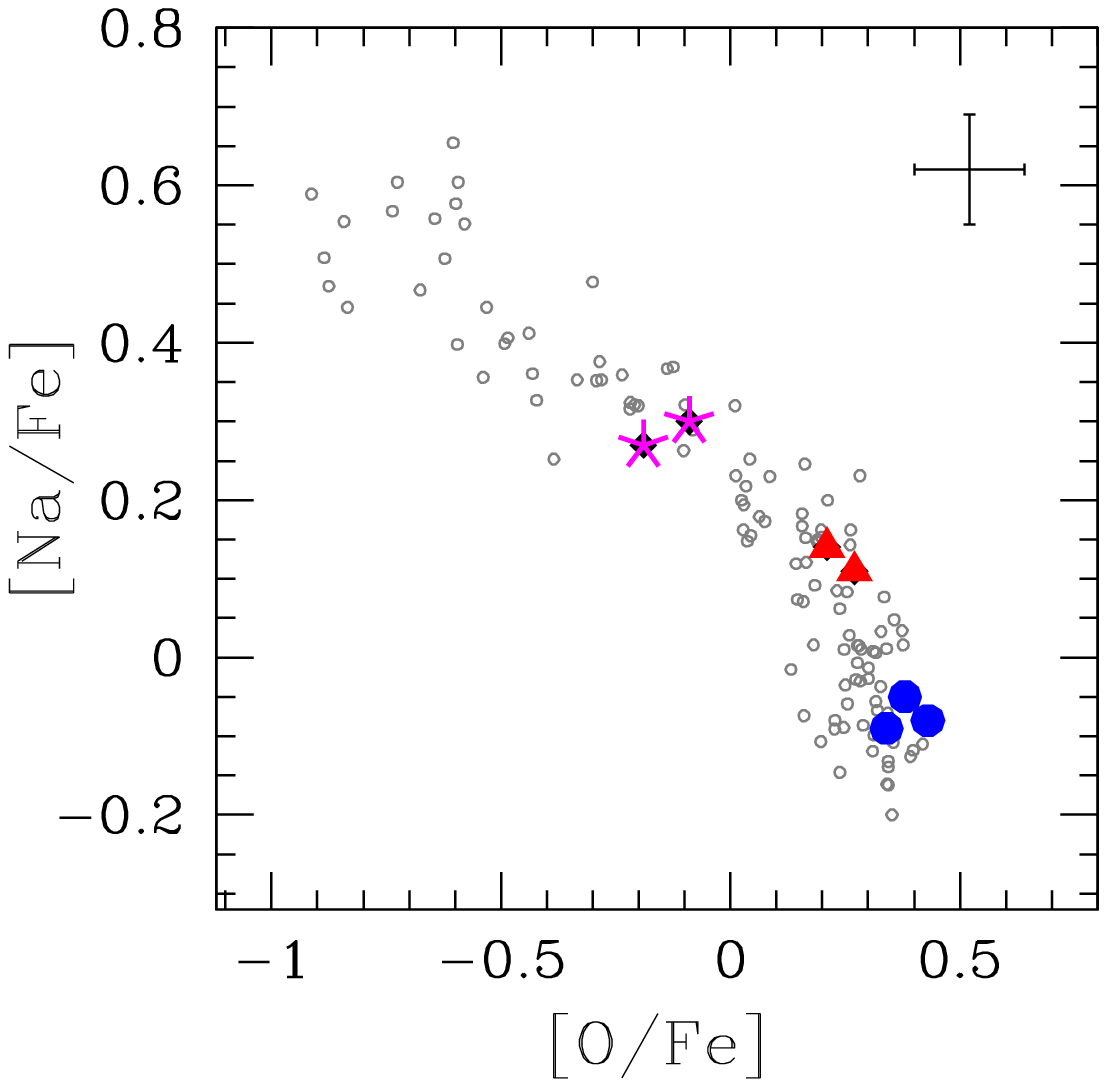}
      \plotone{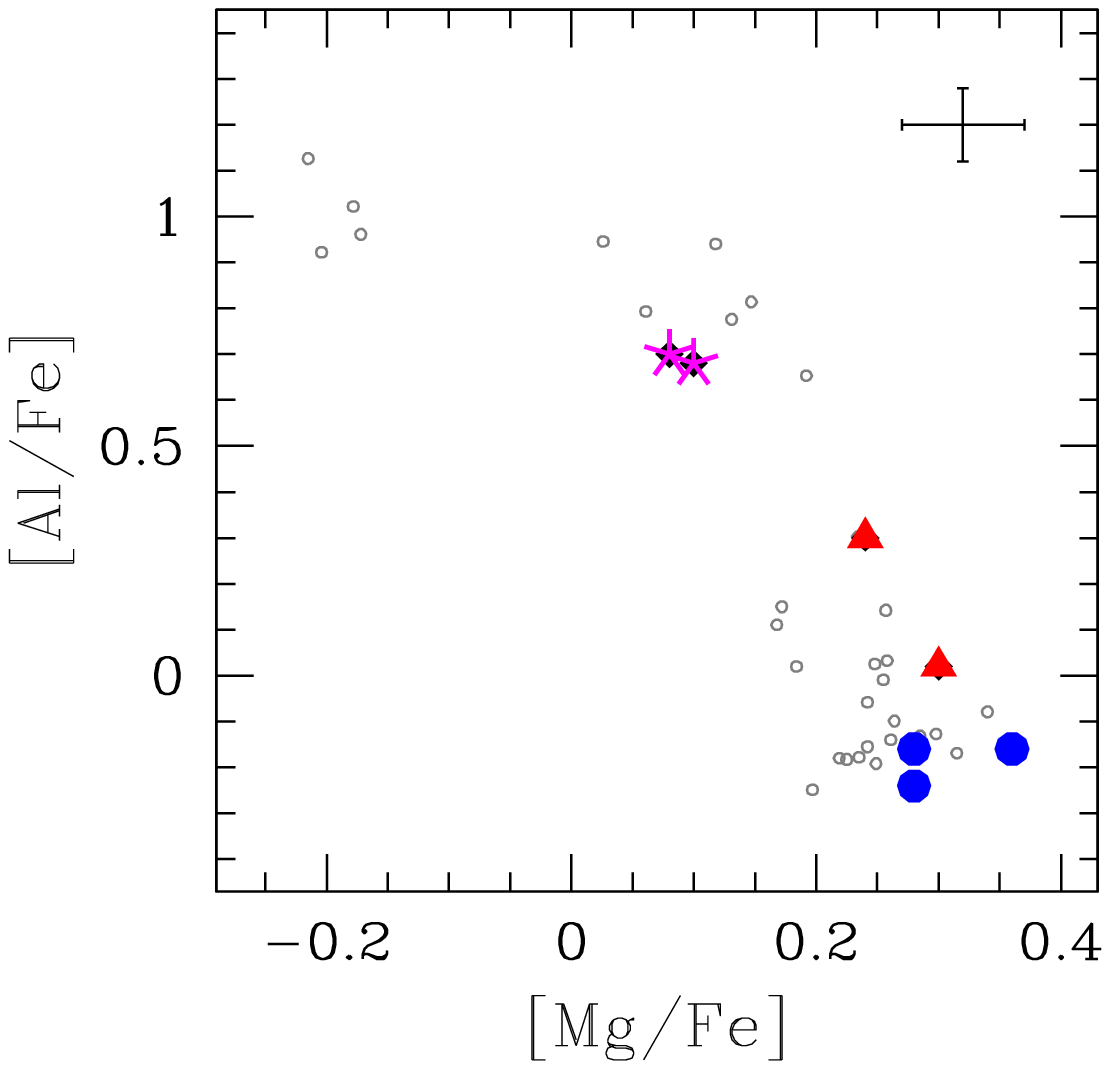}
   \epsscale{.44}
      \plotone{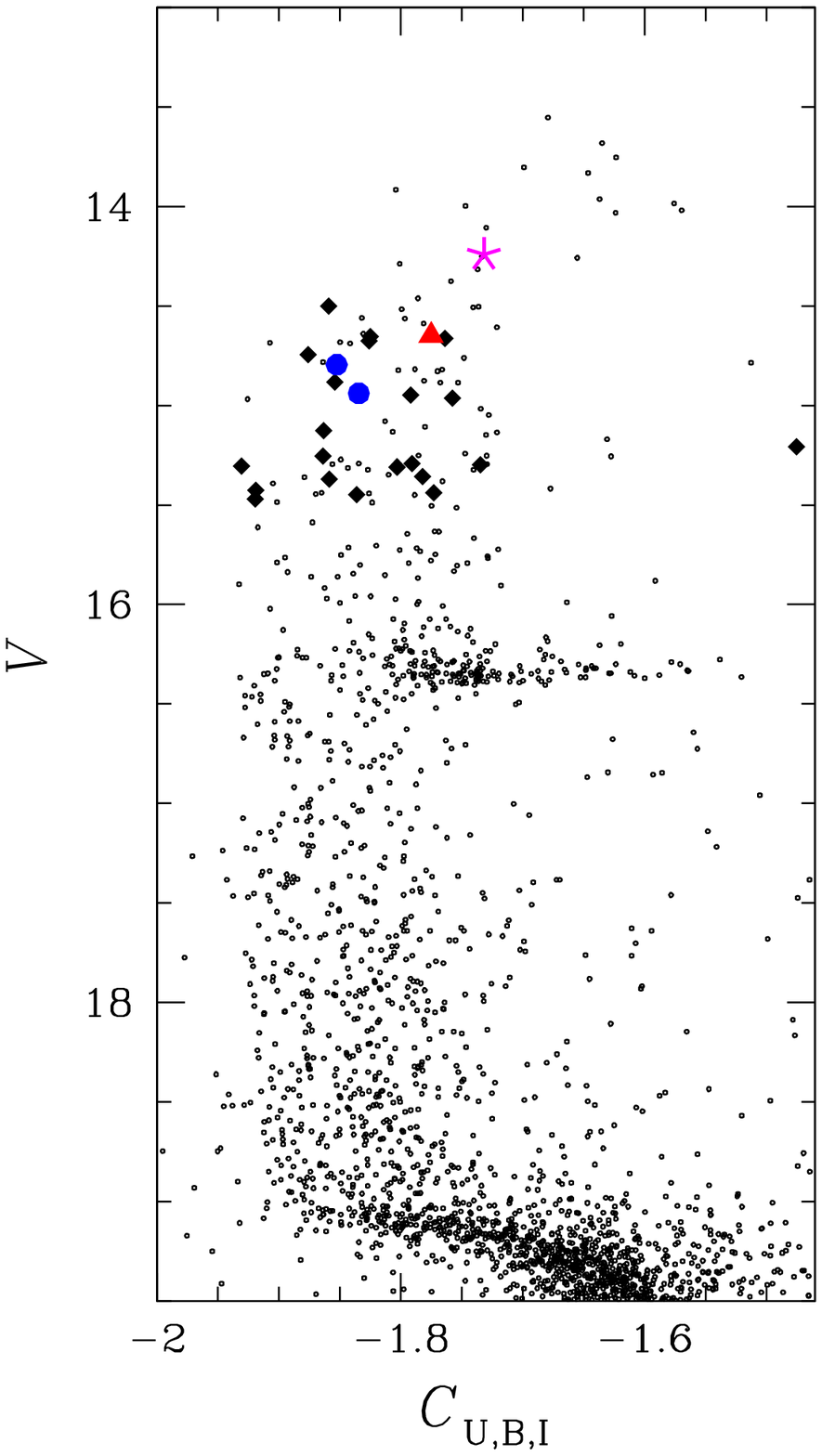}
   \plotone{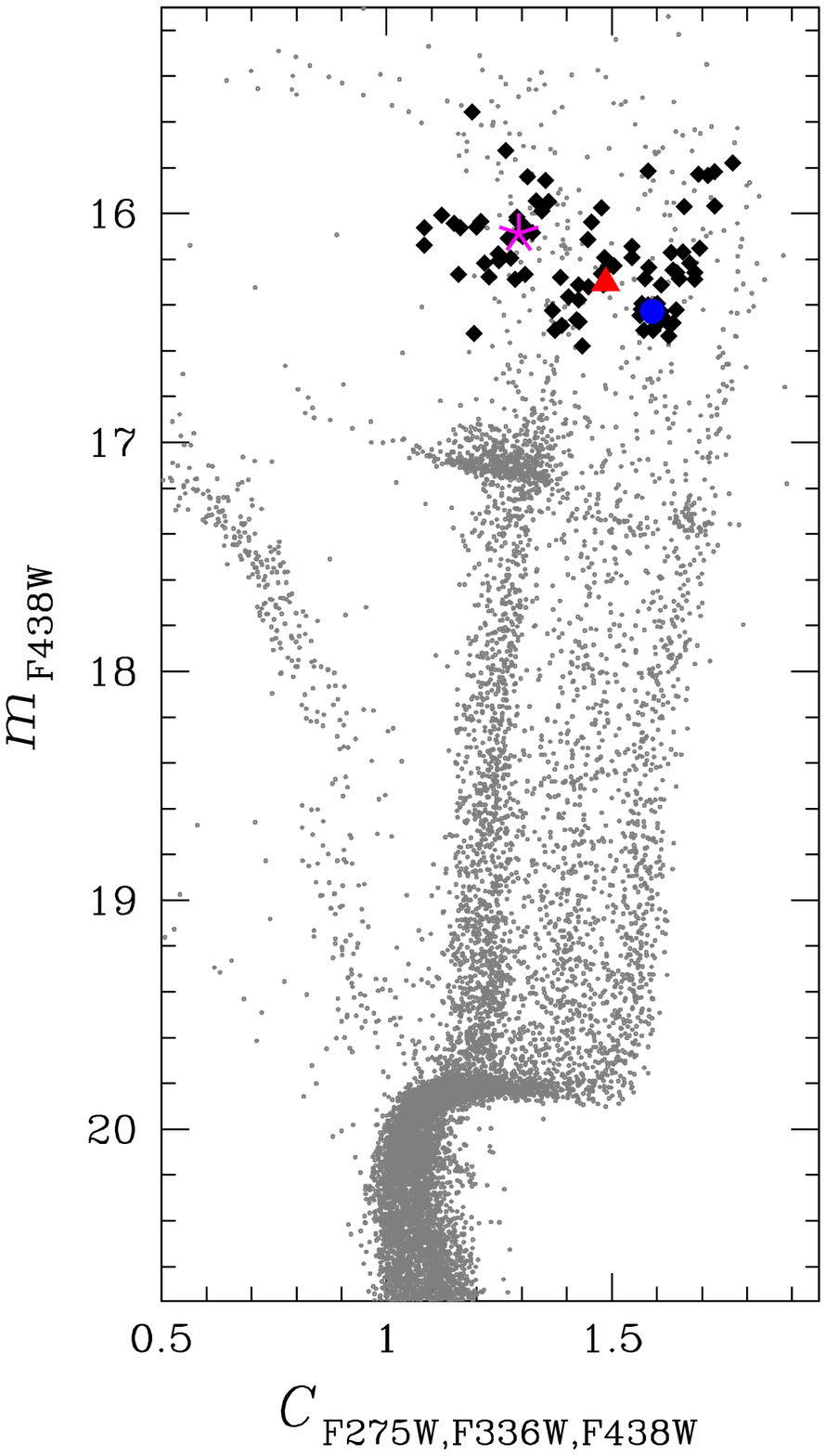}
      \caption{\textit{Upper panels:} Sodium-oxygen and
        magnesium-aluminum anticorrelation for RGB stars (grey
        circles, Carretta\,2014, 2015) and AGB stars (colored
        symbols). \textit{Lower panels:}  $V$ vs.\,$C_{\rm U,B,I}$ (left) and 
        $m_{\rm F438W}$ vs.\,$C_{\rm  F275W,F336W,F438W}$ (right) pseudo-CMDs of NGC\,2808 stars. 
        AGB stars are indicated with black diamonds, while the groups
        1, 2, and 3 of AGB stars (observed spectroscopically),
        selected  on the basis of their position in the [Na/Fe]
        vs.\,[O/Fe] plane, are represented with blue dots, red
        triangles, and magenta starred symbols, respectively. As for
        Figs.~\ref{fig:box} and ~\ref{fig:lightM4}, Al abundances of
        RGB stars have been decreased by 0.21~dex, due to different
        assumed solar abundances. Smaller shifts have been applied
        also to Na and Mg, whose RGB values have been shifted by
        $-$0.05 and $+$0.10~dex, respectively. } 
   \label{fig:light2808}
   \end{figure}

\begin{figure}[ht!]
   \centering
   \epsscale{.8}
   \plotone{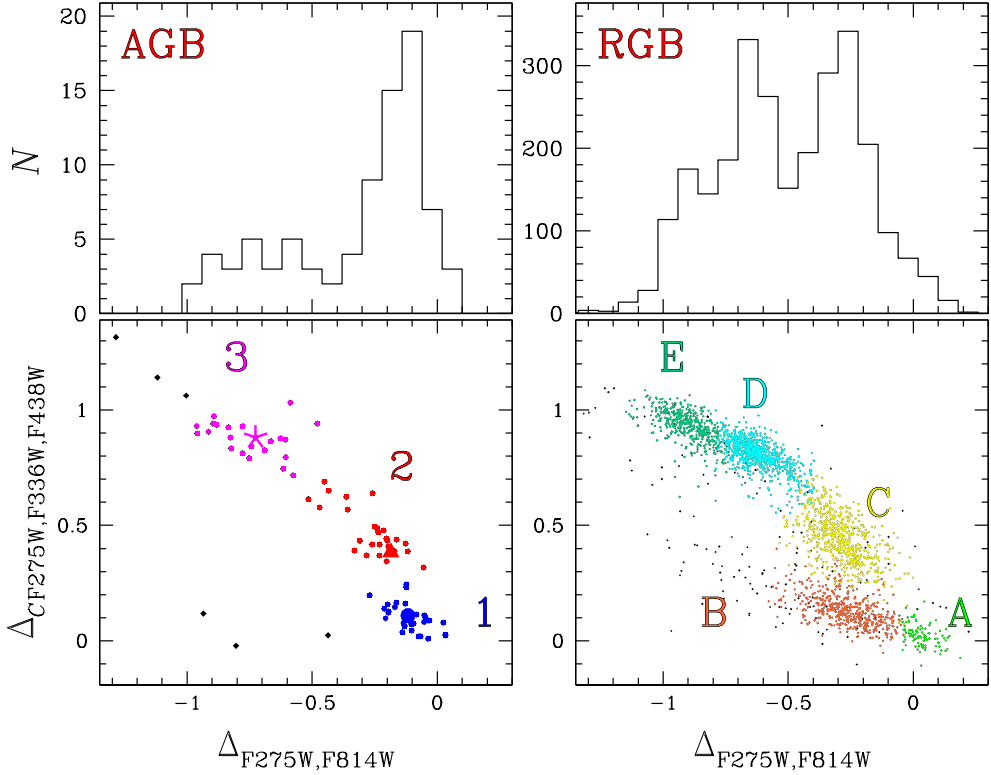}
   \caption{\textit{Upper panels:} $\Delta_{\rm F275W, F814W}$
     histogram distribution for AGB (left) and RGB stars (right) in NGC\,2808.
   \textit{Lower panels:}  $\Delta_{C \rm F275W, F336W, F438W}$
   vs.\,$\Delta_{\rm F275W, F814W}$ pseudo two-color diagram, or
   `chromosome map' of AGB (left) and RGB stars (right) in NGC\,2808
   from Milone et al.\,(2015a). The groups 1, 2, and 3 of AGB stars
   are colored blue, red, and magenta, respectively, while large
   colored symbols indicate our spectroscopic AGB targets. The five
   populations, A--E, of RGB stars are colored green, orange, yellow,
   cyan, and aqua, respectively. Outliers not assigned to any
     population are represented as black points in both the AGB and
     RGB chromosome map.} 
   \label{fig:map2808} 
   \end{figure}

Studies based on high-resolution spectroscopy of RGB stars have revealed
that NGC\,2808 exhibits very extended Na-O and Al-Mg
anticorrelations as shown in the upper panels of
Figure~\ref{fig:light2808}. Here the grey dots represent the [Na/Fe]
vs.\,[O/Fe] and [Al/Fe] vs.\,[Mg/Fe] for RGB stars from Carretta
(2014, 2015). 
Photometry has shown that the MS and RGB of NGC\,2808 host at least
five stellar populations, namely A--E, identified by means of the ``chromosome
map'' tool able to maximise the separation between stellar populations
with different chemical content.
The ``chromosome map'' of NGC\,2808 RGBs published in Milone et
al.\,(2015a) is reproduced in the lower-right panel of
Figure~\ref{fig:map2808}, with the different populations represented
in different colors. 
Populations A--E have indeed distinct combinations of
light-elements/helium abundances (Milone et al.\,2015a):
{\it (i)} Populations E and D
have very high helium content ($Y \sim 0.38$ and $Y \sim 0.32$,
respectively) and extreme abundances of N, O, Na, Mg, and Al. They are
clearly separated from the remaining stars in both the [Na/Fe]-[O/Fe]
and the [Al-Fe]-[Mg-Fe] plane and correspond to the groups of stars
with [O/Fe]$<-$0.2 and $-0.2<$[Mg/Fe]$<0.2$ shown in
Figure~\ref{fig:light2808}. 
{\it (ii)} Stellar populations B and C, which have low sodium and aluminum
 content, are not clearly distinguishable in the diagrams of
 Figure~\ref{fig:light2808}. 
{\it (iii)} There is no spectroscopic information on
 population-A stars, which, according to multi-wavelength photometry
 have similar light-element abundance as populations B and C. 

In this Section we compare the chemical abundances and the photometry
of AGB stars with results on multiple RGBs from the literature and
attempt to connect the populations along the AGB, the RGB, and the MS.
In the upper panels of Fig.~\ref{fig:light2808} we compare our Na-O
and Al-Mg anticorrelations obtained from the AGB stars with those on
the RGB.
This comparison clearly reveals that, among the analyzed AGB stars,
[Na/Fe] anticorrelates with [O/Fe] while [Al/Fe] anticorrelates with
[Mg/Fe], qualitatively confirming that the AGB exhibits the same
chemical pattern as the RGB.
The obvious difference between AGBs and RGBs is that the AGBs do not
reach extremely O-poor and Mg-poor abundances, as RGBs do.

Comparing with the chemical composition observed on the ``chromosome
map'' by Milone et al.\,(2015a), the two AGBs with [O/Fe]$<0.0$ have
chemical composition consistent 
with population-D RGB stars and are represented with magenta starred
symbols in Figure~\ref{fig:light2808}. 
 Two out of the five AGB stars with [O/Fe]$>0.0$ have slightly-higher
 sodium and aluminum abundance than the remaining AGB stars. They
 share the same chemical composition as population-C RGB stars and are
 represented with red triangles. The blue circles represent the
 remaining three stars with [Na/Fe]$<$0.0, which have similar chemical
 abundance as population-B RGB stars.  
For simplicity, in the following, we will refer to the groups of stars
colored blue, red, and magenta, as groups 1, 2, and 3. 
 
The most extreme RGB stars in terms of chemical composition,
e.g. those with the highest He and lowest O, are those belonging to
population-E on the ``chromosome map''.
 None of the seven analyzed AGB stars belong to this group. This fact
 can be either due to the small number of studied AGB stars or to the
 lack of stars with extreme chemical composition along the AGB. 
  By assuming that the population E includes 15\% of the total number
  of NGC\,2808 stars (e.g.\,Simioni et al.\,2016) the probability that
  the lack of AGB population-E stars is due to the small statistical
  sample, inferred from Monte-Carlo simulations, is 0.32. Therefore,
  without additional AGB stars  
  spectroscopically analyzed we cannot draw a firm
  conclusion. By accounting for the radial distribution of different
  stellar populations, we know that population-E stars are more-centrally concentrated
    than the other stars of NGC\,2808, and the fraction of
    population-E stars with respect to the total number of cluster
    stars ranges from 21$\pm$3\% for radial distance $R<$0.6~arcmin
    to 9$\pm$5\% for $R>5.5$~arcmin (Simioni et al.\,2016).  By
    assuming these extreme values for the fraction of population-E
    stars we find that the probability that the lack of AGB
    population-E stars is due to the small statistical sample is 17\%
    and 50\%, respectively. These numbers support the previous
    conclusion that due to the small number of analyzed stars, we
    cannot draw any strong conclusion about the lack of population-E
    stars along the AGB from spectroscopy only. Nevertheless,
    population-E RGB stars are likely to evolve to become the hottest HB
    stars which will then fail to reach the AGB. So, even if the
    statistics is poor, the evidence is consistent with the posit that
    population-E does not reach the AGB.   
    
Hence, 
to further investigate multiple populations along the AGB of
NGC\,2808 we combine information from both spectroscopy and photometry.
   As shown in the lower-right panel of Figure~\ref{fig:light2808},
   the AGB of NGC\,2808 splits into three distinct sequences in the
   $m_{\rm F438W}$ vs.\,$C_{\rm F275W,F336W,F438W}$ pseudo-CMD. 
   We have analyzed the spectra of one star in each of the three
   sequences and find that the three stars belong to the three
   different groups 1, 2, and 3, as previously defined. Therefore they
   have distinct light-element content with the Na abundance
   increasing from the lowest to the highest value of $C_{\rm
     F275W,F336W,F438W}$.  

   Similarly to the AGB, the RGB of NGC\,2808 exhibits three main
   sequences in the $m_{\rm F438W}$ vs.\,$C_{\rm F275W,F336W,F438W}$
   pseudo-CMD  (Piotto et al.\,2015; Milone et al.\,2015a). 
   Most of the difference in the $C_{\rm F275W,F336W,F438W}$
   pseudo-color among stars with the same magnitude is interpreted as
   the effect of light-element variation in the spectrum of the star. 
   Thus, we expect that stars in the three AGB photometric sequences, shown in
   the lower-right panel of Figure~\ref{fig:light2808}, are associated with
   the three corresponding main RGBs. 
   
The fact that the AGB of NGC\,2808 is not consistent with a simple
population is further supported by the $V$ vs. $C_{\rm U,B,I}$
pseudo-CMD from ground-based photometry plotted in the lower-left
panel of Figure~\ref{fig:light2808}, where the $C_{\rm U,B,I}$
dispersion of AGB stars is significantly larger than that expected
from observational errors only. 
In contrast with the RGB, where there is a clear correlation
(anti-correlation) between the $C_{\rm U,B,I}$ value of a star and its
sodium (oxygen) abundance (Monelli et al.\,2013), the small number of
analyzed AGB stars prevents us from any strong conclusion about the
possible correlation between the light-element abundance of an AGB
star and its $C_{\rm U,B,I}$ pseudo-color. Nevertheless, we note that
Na-poor AGB stars have smaller $C_{\rm U,B,I}$ values than Na-rich AGB
stars, in close analogy with what we have observed in
M\,4. \footnote{We note that the two stellar populations of AGB
  and RGB stars in M\,4 distribute along two distinct sequences in the $V$
  vs. $C_{\rm U,B,I}$ diagram, in contrast with what we observe in
  NGC\,2808 where it is impossible to distinguish the different
  stellar populations along the RGB and AGB by using the $C_{\rm
    U,B,I}$ index only (see Monelli et al.\,2013 for the $C_{\rm
    U,B,I}$ distribution of RGB stars). The position of a star in the
  $V$ vs. $C_{\rm U,B,I}$ pseudo-CMD depends on its abundance of
  helium and light elements and accurate study involving isochrones
  and synthetic spectra  with appropriate chemical composition is
  mandatory to fully understand the observed distribution of stars in
  the $V$ vs. $C_{\rm U,B,I}$ (see e.g.\,Dotter et al.\,2015 for
  NGC\,6752). Similar conclusion can be extended to the $m_{\rm
    F438W}$ vs.\,$C_{\rm F275W,F336W,F438W}$ diagram.}

In the lower-left panel of Figure~\ref{fig:map2808}, we plotted the
$\Delta_{C \rm F275W, F336W, F438W}$ vs.\,$\Delta_{\rm F275W, F814W}$
pseudo-CMD or `chromosome' map of AGB stars in NGC\,2808 that we have
derived by extending to AGB stars the method that we have previously
introduced for the RGB (Milone et al.\,2015a). The lower-right panel of
the same figure reproduces the same diagram derived for RGB stars and
marks the main populations A--E of NGC\,2808 with different colors. 
 In contrast to the observations of MS and RGB stars, where at least
 five stellar populations are present in the chromosome map, the
 lower-left panel of Figure~\ref{fig:map2808} reveals only three
 groups of AGB stars.
 For example, relative to the number of population D stars on the AGB,
 there are many fewer stars in the population E region.  Similarly,
 there is an apparent dearth of AGB stars in the population A region. 

We followed the recipe by McLachlan \& Peel (2000) to derive the
groups of AGB stars that are statistically significant. Briefly, we
determined the maximum-likelihood fit to various numbers of groups and
calculated the optimal number of groups by  using the Bayesian
information criterion (BIC) penalized likelihood measure for model
complexity.
To this aim, we did vary the size and the shape of the distinct
  groups of AGB stars. For each combination of shape and size, we
  assumed a number, $N$ of populations from 1 to 8, and calculated a
  BIC value. The most likely explanation corresponds to BIC=79.2 and
  $N=4$ with the assumption that the groups have equal shape and
  variable volume and orientations (VEV). Similarly, the second best
  BIC value (BIC=74.2) corresponds to $N=4$ and stellar groups with
  variable shape, volume and orientations (VVV). The resulting three
  main groups of stars are colored in blue, red, magenta, in
  Fig.~\ref{fig:map2808}. The fourth stellar group includes six
  only outliers that have been colored black. 
 The results support the visual impression that the AGB of
NGC\,2808 hosts only three main stellar, strengthening the idea that
stars with extreme He and O do not evolve through the AGB phase. 
 
We note that the ``chromosome maps'' of RGB and AGB stars reveal significant differences.
  Specifically, only one group of stars with $\Delta_{\rm
    F275W,F336W,F438W} \sim 0.0$ is present along the AGB, in contrast
  with what we observe along the RGB where we clearly distinguish the
  two groups of population-A and population-B stars. 
  It remains unclear whether population-A stars do not exist along the
  AGB or if they are mixed with population-B. 
  
  Moreover, while we clearly observe population-D and -E stars along
  the RGB and the MS, only one group of AGB stars with large values
  $\Delta_{\rm F275W,F336W,F438W}$ is present. To investigate the
  presence of population-E AGB stars in    
 the upper panels of  Figure~\ref{fig:map2808} we compare the
 $\Delta_{\rm F275W, F814W}$ histogram distribution of AGB stars and
 RGB stars. As discussed in Milone et al.\,(2015a, see their Fig.~3)
 the RGB of NGC\,2808 is consistent with three main peaks that
 correspond to population E, D, and to the group of populations
 A$+$B$+$C. In contrast, along the AGB we distinguish a dominant peak,
 associated to the stellar groups 1 and 2, plus a stellar tail mostly
 due to group-3 stars. 
 
 If group-3 includes both populations D and E the fraction of group-3
 stars with respect to the total number of AGB stars should be
 consistent with the ratio between population D$+$E RGB stars and the
 total number of RGB stars. 
 We find that group 3 includes 27$\pm$5\% of the total number of
 AGB. This value significantly differs from the ratio of RGB-D$+$E
 stars with respect to the total number of AGB stars that is
 50$\pm$1\%. In contrast, the ratio between RGB-D stars and the number
 of RGB-A$+$B$+$C$+$D star is 30$\pm$1\% and is consistent within one
 sigma with the fraction of group-3 AGB stars with respect to the
 total number of AGB stars. 
These results are consistent with the lack of population-E stars along the AGB.

%__________________________________________________________________
%__________________________________________________________________

 %__________________________________________________________________
%__________________________________________________________________

%%%%%%%%%%%%%%%%%%%%%%%%%%%%%%%%%%%%%%%%%%%%%%%%%%%%
\begin{landscape}
\begin{table}
  \center
  \scriptsize
\begin{tabular}{ccccccccccc}
\hline
\hline
ID (2MASS)       &  GC        & RA           & DEC            & RV & \teff & \logg & [Fe/H] & \vmicro & \teff (phot) & \logg (phot) \\
                 &            & J2000        & J2000          &[\kmsec]& [K]&[cgs] &        & [\kmsec]& [K]          & [cgs]   \\
\hline
16233067-2629390 & M\,4       & 16:23:30.70  & $-$26:29:39.0   &  $+$72.88     & 4920 & 1.85 & $-$1.18 & 1.81 & 4817 & 1.77\\ %tu9
16233741-2638238 & M\,4       & 16:23:37.44  & $-$26:38:23.9   &  $+$66.88     & 4770 & 1.50 & $-$1.22 & 1.72 & 4732 & 1.58\\ %tu16
16234268-2631209 & M\,4       & 16:23:42.71  & $-$26:31:20.8   &  $+$74.01     & 5190 & 2.25 & $-$1.07 & 1.78 & 5022 & 1.92\\ %tu19
16233020-2633241 & M\,4       & 16:23:30.23  & $-$26:33:24.0   &  $+$67.26     & 4630 & 1.55 & $-$1.19 & 1.91 & 4641 & 1.44\\ %tu8
16235035-2632478 & M\,4       & 16:23:50.38  & $-$26:32:47.8   &  $-$48.01     &  --  &  --  &   --    &  --  &  --  &  -- \\ %tu21
16233142-2633110 & M\,4       & 16:23:31.45  & $-$26:33:10.9   &  $+$65.80     & 4400 & 1.25 & $-$1.18 & 2.00 & 4445 & 1.18\\ %tu10
16233846-2629235 & M\,4       & 16:23:38.49  & $-$26:29:23.5   &  $+$68.32     & 4780 & 1.75 & $-$1.17 & 1.82 & 4701 & 1.63\\ %tu17
16233535-2632225 & M\,4       & 16:23:35.37  & $-$26:32:22.5   &  $+$67.83     & 4450 & 1.20 & $-$1.26 & 1.92 & 4497 & 1.27\\ %tu13
16240858-2624552 & M\,4       & 16:24:08.60  & $-$26:24:55.2   &  $+$72.82     & 4680 & 1.60 & $-$1.27 & 1.82 & 4623 & 1.46\\ %tu25
16233477-2631349 & M\,4       & 16:23:34.79  & $-$26:31:34.9   &  $+$69.68     & 5150 & 2.50 & $-$1.22 & 1.62 & 5023 & 1.92\\ %tu12
16234740-2631463 & M\,4       & 16:23:47.41  & $-$26:31:46.3   &  $+$72.98     & 4500 & 1.35 & $-$1.28 & 1.92 & 4536 & 1.32\\ %tu20
16232988-2631490 & M\,4       & 16:23:29.90  & $-$26:31:49.0   &  $+$71.86     & 4530 & 1.40 & $-$1.28 & 1.77 & 4561 & 1.39\\ %tu7
16235375-2634426 & M\,4       & 16:23:53.77  & $-$26:34:42.6   &  $+$65.54     & 4530 & 1.40 & $-$1.21 & 1.93 & 4547 & 1.31\\ %tu22
16233667-2630397 & M\,4       & 16:23:36.69  & $-$26:30:39.7   &  $+$63.13     & 4470 & 1.23 & $-$1.24 & 1.92 & 4508 & 1.28\\ %tu15
16233614-2632015 & M\,4       & 16:23:36.17  & $-$26:32:01.5   &  $+$67.30     & 4570 & 1.75 & $-$1.12 & 1.57 & 4570 & 1.31\\ %tu14
16231672-2634279 & M\,4       & 16:23:16.75  & $-$26:34:28.0   &  $+$72.19     & 4500 & 1.40 & $-$1.24 & 1.96 & 4435 & 1.20\\ %tu3
16234085-2631215 & M\,4       & 16:23:40.88  & $-$26:31:21.5   &  $+$71.45     & 4550 & 1.60 & $-$1.18 & 1.70 & 4587 & 1.48\\ %tu18
16232114-2631598 & M\,4       & 16:23:21.17  & $-$26:31:59.7   &  $+$71.96     & 5090 & 2.47 & $-$1.14 & 1.61 & 4964 & 2.00\\ %tu4
09120251-6451001 & NGC\,2808  & 09:12:02.51  & $-$64:51:00.2   &  $+$96.61     & 4750 & 1.55 & $-$1.31 & 1.70 & 4773 & 1.66\\ %2_43a
09123016-6454129 & NGC\,2808  & 09:12:30.18  & $-$64:54:12.9   &  $+$99.14     & 4770 & 1.78 & $-$1.20 & 1.55 & 4796 & 1.54\\ %2_269
09120852-6449107 & NGC\,2808  & 09:12:08.53  & $-$64:49:10.7   & $+$110.92     & 4630 & 1.55 & $-$1.19 & 1.75 & 4690 & 1.55\\ %2_168
09120213-6452243 & NGC\,2808  & 09:12:02.15  & $-$64:52:24.4   &  $+$87.84     & 4860 & 1.55 & $-$1.30 & 1.82 & 4788 & 1.62\\ %3_41a
09122027-6448450 & NGC\,2808  & 09:12:20.29  & $-$64:48:45.1   &  $+$86.07     & 4500 & 1.15 & $-$1.32 & 1.75 & 4579 & 1.36\\ %2_229
09120665-6450253 & NGC\,2808  & 09:12:06.66  & $-$64:50:25.4   & $+$106.72     & 4800 & 1.60 & $-$1.30 & 1.70 & 4887 & 1.80\\ %1_45a
09114655-6452144 & NGC\,2808  & 09:11:46.55  & $-$64:52:14.5   &  $+$89.63     & 4400 & 0.90 & $-$1.25 & 2.05 & 4420 & 1.07\\ %2_63
\hline
\end{tabular}
\label{tab:data}
\caption{Coordinates, radial velocities, and atmospheric parameters of
  the AGB stars spectroscopically analysed in this paper.}
\end{table}
\end{landscape}

%%%%%%%%%%%%%%%%%%%%%%%%%%%%%%%%%%%%%%%%%%%%%%%%%%%%
\begin{table*}
\caption{Sensitivity of derived abundances to the uncertainties in atmospheric parameters, the limited S/N ($\sigma_{\rm fit}$) and the total error due to these contributions ($\sigma_{\rm tot}$).\label{tab:err}}
\begin{tabular}{lcccccc}
\hline\hline
                       &$\Delta$\teff      &$\Delta$\logg      &$\Delta$\vmicro    & $\Delta$[m/H]      & $\sigma_{\rm fit}$  &  $\sigma_{\rm total}$\\  
                       &$\pm$100~K         & $\pm$0.20         &$\pm$0.30~\kmsec   &    0.15~dex        &                    &                     \\
\hline
$\rm {[O/Fe]}$           &$\pm$0.02        &$\pm$0.09          &$\mp$0.01          &$\mp$0.05           &$\pm$0.06    & 0.12 \\  % sulla singola riga l'err sul continuo e' circa 0.08. La somma in quadratura per 2 righe e' 0.06
$\rm {[Na/Fe]}$          &$\mp$0.02        &$\mp$0.01          &$\pm$0.03          &$\mp$0.01           &$\pm$0.06    & 0.07 \\
$\rm {[Mg/Fe]}$          &$\mp$0.03        &$\mp$0.01          &$\pm$0.00          &$\mp$0.00           &$\pm$0.04    & 0.05 \\
$\rm {[Al/Fe]}$          &$\pm$0.07        &$\mp$0.00          &$\mp$0.00          &$\pm$0.01           &$\pm$0.04    & 0.08 \\  % sulla singola riga l'err sul continuo e' circa 0.08. La somma in quadratura per 2 righe e' 0.06
$\rm {[Si/Fe]}$          &$\mp$0.08        &$\pm$0.02          &$\pm$0.04          &$\pm$0.02           &$\pm$0.07    & 0.12 \\
$\rm {[Ca/Fe]}$          &$\pm$0.01        &$\mp$0.02          &$\mp$0.04          &$\mp$0.01           &$\pm$0.01    & 0.05 \\
$\rm {[Sc/Fe]}$\,{\sc i} &$\pm$0.00        &$\pm$0.01          &$\pm$0.04          &$\pm$0.01           &$\pm$0.03    & 0.05 \\
$\rm {[Sc/Fe]}$\,{\sc ii}&$\pm$0.05        &$\mp$0.01          &$\mp$0.03          &$\pm$0.01           &$\pm$0.05    & 0.08 \\
$\rm {[Ti/Fe]}$\,{\sc i} &$\pm$0.06        &$\mp$0.01          &$\mp$0.01          &$\mp$0.01           &$\pm$0.01    & 0.06 \\
$\rm {[Ti/Fe]}$\,{\sc ii}&$\pm$0.04        &$\mp$0.01          &$\mp$0.06          &$\mp$0.01           &$\pm$0.03    & 0.08 \\
$\rm {[V/Fe]}$           &$\pm$0.07        &$\mp$0.01          &$\pm$0.02          &$\mp$0.01           &$\pm$0.03    & 0.08 \\
$\rm {[Cr/Fe]}$\,{\sc i} &$\pm$0.04        &$\mp$0.01          &$\mp$0.02          &$\mp$0.01           &$\pm$0.02    & 0.05 \\
$\rm {[Cr/Fe]}$\,{\sc ii}&$\pm$0.01        &$\mp$0.01          &$\pm$0.02          &$\mp$0.01           &$\pm$0.04    & 0.05 \\
$\rm {[Fe/H]}$\,{\sc i}  &$\pm$0.10        &$\pm$0.01          &$\mp$0.06          &$\pm$0.00           &$\pm$0.01    & 0.12 \\
$\rm {[Fe/H]}$\,{\sc ii} &$\mp$0.06        &$\pm$0.09          &$\mp$0.03          &$\pm$0.04           &$\pm$0.04    & 0.13 \\
$\rm {[Co/Fe]}$          &$\mp$0.01        &$\pm$0.00          &$\pm$0.05          &$\pm$0.01           &$\pm$0.04    & 0.07 \\
$\rm {[Ni/Fe]}$          &$\mp$0.02        &$\pm$0.01          &$\pm$0.03          &$\pm$0.01           &$\pm$0.01    & 0.04 \\
$\rm {[Zn/Fe]}$          &$\mp$0.13        &$\pm$0.04          &$\mp$0.07          &$\pm$0.02           &$\pm$0.02    & 0.16 \\
$\rm {[Y/Fe]}$\,{\sc ii} &$\pm$0.06        &$\mp$0.01          &$\mp$0.09          &$\mp$0.00           &$\pm$0.05    & 0.12 \\
$\rm {[Ce/Fe]}$\,{\sc ii}&$\pm$0.08        &$\mp$0.01          &$\pm$0.01          &$\pm$0.01           &$\pm$0.07    & 0.11 \\
\hline
\end{tabular}
\label{tab1}
\end{table*} 

%%%%%%%%%%%%%%%%%%%%%%%%%%%%%%%%%%%%%%%%%%%%%%%%%%%%
\begin{landscape}
\begin{table}
\center
\caption{Derived chemical abundances for O, Na, Mg, Al, Si, and Ca in M\,4 and NGC\,2808 AGBs.
  \label{tab:abu1}}
        %{
          \scriptsize
\begin{tabular}{ ccc ccc ccc ccc c}
\\
  \hline
\hline
ID (2MASS) & $\rm {[O/Fe]}$ & rms/\#  & $\rm {[Na/Fe]}$ & rms/\#  & $\rm {[Mg/Fe]}$ & rms/\#  & $\rm {[Al/Fe]}$ & rms/\#  & $\rm {[Si/Fe]}$ & rms/\#  & $\rm {[Ca/Fe]}$ & rms/\# \\
\hline
\multicolumn{13}{c}{M\,4} \\
    16233142-2633110   &  $+$0.34 & 0.04/2 & $+$0.25 & 0.13/3 & $+$0.51 & 0.07/3 & $+$0.34 & 0.04/2 & $+$0.49 & 0.01/3 & $+$0.19 & 0.09/13\\%tu10
    16233477-2631349   &  $+$0.57 & 0.02/2 & $+$0.09 & 0.12/3 & $+$0.46 & 0.09/2 & $+$0.27 & 0.06/2 & $+$0.42 & 0.04/3 & $+$0.32 & 0.09/14\\%tu12
    16233535-2632225   &  $+$0.34 & 0.04/2 & $+$0.37 & 0.09/3 & $+$0.53 & 0.00/2 & $+$0.38 & 0.03/2 & $+$0.51 & 0.04/3 & $+$0.20 & 0.08/13\\
    16233614-2632015   &  $+$0.56 & 0.02/2 & $+$0.24 & 0.07/3 & $+$0.44 & 0.02/3 & $+$0.23 & 0.00/2 & $+$0.42 & 0.06/3 & $+$0.31 & 0.09/13\\%tu14
    16233741-2638238   &  $+$0.24 & 0.04/2 & $+$0.27 & 0.14/3 & $+$0.48 & 0.09/2 & $+$0.28 & 0.02/2 & $+$0.46 & 0.01/3 & $+$0.21 & 0.11/12\\%tu16
    16234085-2631215   &  $+$0.53 & 0.01/2 & $+$0.02 & 0.06/3 & $+$0.61 & 0.00/2 & $+$0.50 & 0.05/2 & $+$0.56 & 0.02/3 & $+$0.21 & 0.10/13\\%tu18
    16234268-2631209   &  $+$0.40 & 0.01/2 & $+$0.32 & 0.16/4 & $+$0.35 & --  /1 & $+$0.25 & 0.04/2 & $+$0.47 & 0.02/3 & $+$0.30 & 0.10/11\\%tu19
    16234740-2631463   &  $+$0.53 & 0.03/2 & $+$0.09 & 0.06/3 & $+$0.54 & 0.02/3 & $+$0.31 & 0.02/2 & $+$0.51 & 0.04/3 & $+$0.18 & 0.09/13\\%tu20
    16235375-2634426   &  $+$0.37 & 0.04/2 & $+$0.36 & 0.08/3 & $+$0.49 & 0.04/3 & $+$0.31 & 0.03/2 & $+$0.53 & 0.06/2 & $+$0.18 & 0.10/13\\%tu22
    16240858-2624552   &  $+$0.58 & 0.02/2 & $-$0.07 & 0.10/3 & $+$0.47 & 0.06/2 & $+$0.20 & 0.04/2 & $+$0.45 & 0.07/3 & $+$0.17 & 0.09/13\\
    16231672-2634279   &  $+$0.46 & 0.01/2 & $+$0.29 & 0.06/3 & $+$0.51 & 0.02/2 & $+$0.30 & 0.01/2 & $+$0.51 & 0.05/3 & $+$0.17 & 0.10/13\\%tu3
    16232114-2631598   &  $+$0.54 & 0.13/2 & $+$0.13 & 0.07/4 & $+$0.45 & 0.06/2 & $+$0.19 & 0.04/2 & $+$0.46 & 0.07/3 & $+$0.30 & 0.08/13\\
    16232988-2631490   &  $+$0.50 & 0.01/2 & $+$0.02 & 0.08/3 & $+$0.55 & 0.01/2 & $+$0.32 & 0.04/2 & $+$0.51 & 0.03/3 & $+$0.19 & 0.09/13\\%tu7
    16233020-2633241   &  $+$0.35 & 0.04/2 & $+$0.36 & 0.19/4 & $+$0.51 & 0.05/2 & $+$0.34 & 0.04/2 & $+$0.52 & 0.03/3 & $+$0.20 & 0.10/12\\%tu8
    16233067-2629390   &  $+$0.50 & 0.06/2 & $+$0.00 & 0.13/4 & $+$0.52 & 0.22/2 & $+$0.17 & 0.01/2 & $+$0.52 & 0.02/2 & $+$0.22 & 0.11/ 9\\%tu9
    16233667-2630397   &  $+$0.36 & 0.01/2 & $+$0.34 & 0.06/3 & $+$0.52 & 0.01/3 & $+$0.35 & 0.01/2 & $+$0.51 & 0.04/3 & $+$0.19 & 0.09/12\\
    16233846-2629235   &  $+$0.44 & 0.04/2 & $+$0.17 & 0.09/3 & $+$0.48 & 0.06/2 & $+$0.23 & 0.01/2 & $+$0.46 & 0.05/3 & $+$0.20 & 0.09/12\\%tu17
\hline                                                  
 mean                  & $+$0.45  &        & $+$0.19 &        & $+$0.50 &        & $+$0.29 &        & $+$0.49 &        & $+$0.22 &        \\              
 $\pm$                 & 0.02     &        & 0.04    &        & 0.01    &        & 0.02    &        & 0.01    &        & 0.01    &        \\              
 rms                   & 0.10     &        & 0.14    &        & 0.05    &        & 0.08    &        & 0.04    &        & 0.04    &        \\              
\hline
\multicolumn{13}{c}{NGC\,2808} \\
09120665-6450253    &  $+$0.38 & 0.03/2 & $-$0.05 & 0.07/4 & $+$0.28 & 0.01/2 & $-$0.16 & 0.00/2 & $+$0.32 & 0.09/3 & $+$0.26 & 0.09/14\\
09120251-6451001    &  $+$0.27 & 0.07/2 & $+$0.11 & 0.11/4 & $+$0.24 & 0.04/2 & $+$0.30 & 0.04/2 & $+$0.23 & 0.02/3 & $+$0.22 & 0.10/11\\
09120213-6452243    &  $-$0.19 & 0.09/2 & $+$0.27 & 0.07/4 & $+$0.08 & 0.14/2 & $+$0.70 & 0.01/2 & $+$0.30 & 0.03/3 & $+$0.17 & 0.13/12\\
09114655-6452144    &  $-$0.09 & 0.08/2 & $+$0.30 & 0.10/4 & $+$0.10 & 0.02/3 & $+$0.68 & 0.03/2 & $+$0.29 & 0.05/3 & $+$0.18 & 0.11/10\\
09122027-6448450    &  $+$0.21 & 0.01/2 & $+$0.14 & 0.09/4 & $+$0.30 & 0.05/3 & $+$0.02 & 0.01/2 & $+$0.25 & 0.06/3 & $+$0.22 & 0.11/13\\
09123016-6454129    &  $+$0.34 & 0.04/2 & $-$0.09 & 0.11/4 & $+$0.28 & 0.01/3 & $-$0.24 & 0.07/2 & $+$0.24 & 0.04/3 & $+$0.31 & 0.10/14\\
09120852-6449107    &  $+$0.43 & 0.04/2 & $-$0.08 & 0.13/4 & $+$0.36 & 0.06/2 & $-$0.16 & 0.04/2 & $+$0.30 & 0.14/2 & $+$0.30 & 0.08/14\\
\hline
 mean                  & $+$0.19  &        & $+$0.09 &        & $+$0.23 &        & $+$0.16 &        & $+$0.28 &        & $+$0.24 &        \\              
 $\pm$                 & 0.09     &        & 0.06    &        & 0.04    &        & 0.15    &        & 0.01    &        & 0.02    &        \\              
 rms                   & 0.22     &        & 0.15    &        & 0.10    &        & 0.37    &        & 0.03    &        & 0.05    &        \\              
\hline
%}
\end{tabular}
\end{table}
\end{landscape}

%%%%%%%%%%%%%%%%%%%%%%%%%%%%%%%%%%%%%%%%%%%%%%%%%%%%%%%%%%%%%%%%%%%%

\begin{landscape}
\begin{table}
\center
\caption{Derived chemical abundances for Sc, Ti, V, and Cr in M\,4 and NGC\,2808 AGBs.
  \label{tab:abu2}}
          \scriptsize
\begin{tabular}{c ccc ccc ccc ccc cc}
\\
  \hline
\hline
ID (2MASS) &$\rm {[Sc/Fe]}$\,{\sc i} & rms/\#  &$\rm {[Sc/Fe]}$\,{\sc ii} & rms/\#  &$\rm {[Ti/Fe]}$\,{\sc i} & rms/\#  &$\rm {[Ti/Fe]}$\,{\sc ii} & rms/\#  &$\rm {[V/Fe]}$ & rms/\#  &$\rm {[Cr/Fe]}$\,{\sc i} & rms/\#  &$\rm {[Cr/Fe]}$\,{\sc ii} & rms/\#\\
\hline
\multicolumn{15}{c}{M\,4} \\

16233142-2633110  & --      & --  /0 & $+$0.01 & 0.11/6 & $+$0.22 & 0.06/15 & $+$0.30 & 0.15/5 & $-$0.12 & 0.30/13 & $-$0.12 & 0.14/10 & $-$0.05 & 0.04/2 \\
16233477-2631349  & $-$0.02 & --  /1 & $+$0.14 & 0.04/6 & $+$0.32 & 0.08/15 & $+$0.36 & 0.08/5 & $+$0.01 & 0.11/10 & $-$0.06 & 0.10/ 7 & $-$0.01 & 0.02/2 \\
16233535-2632225  & $-$0.07 & --  /1 & $-$0.01 & 0.07/5 & $+$0.23 & 0.05/16 & $+$0.21 & 0.06/4 & $-$0.03 & 0.08/12 & $-$0.16 & 0.11/ 7 & $-$0.08 & --  /1 \\
16233614-2632015  & $+$0.04 & --  /1 & $+$0.15 & 0.06/6 & $+$0.29 & 0.06/15 & $+$0.28 & 0.03/4 & $+$0.07 & 0.06/12 & $-$0.04 & 0.10/ 9 & $-$0.05 & --  /1 \\
16233741-2638238  & $-$0.09 & 0.23/6 & $-$0.12 & 0.23/6 & $+$0.22 & 0.09/13 & $+$0.27 & 0.15/5 & $-$0.16 & 0.18/12 & $-$0.13 & 0.10/ 7 & $-$0.23 & --  /1 \\
16234085-2631215  & $-$0.05 & --  /1 & $+$0.02 & 0.06/6 & $+$0.23 & 0.06/14 & $+$0.25 & 0.07/5 & $-$0.05 & 0.04/12 & $-$0.15 & 0.14/ 6 & $-$0.16 & --  /1 \\
16234268-2631209  & $-$0.01 & 0.22/5 & $-$0.04 & 0.22/5 & $+$0.30 & 0.07/10 & $+$0.30 & 0.10/5 & $-$0.01 & 0.10/ 4 & $-$0.10 & 0.09/ 6 & $-$0.09 & --  /1 \\
16234740-2631463  & $-$0.04 & --  /1 & $-$0.01 & 0.06/6 & $+$0.25 & 0.04/13 & $+$0.27 & 0.12/5 & $-$0.01 & 0.08/13 & $-$0.15 & 0.10/ 7 & $-$0.12 & --  /1 \\
16235375-2634426  & $-$0.12 & --  /1 & $+$0.04 & 0.04/6 & $+$0.23 & 0.05/16 & $+$0.25 & 0.11/5 & $-$0.01 & 0.08/13 & $-$0.15 & 0.10/ 8 & $-$0.09 & --  /1 \\
16240858-2624552  & $-$0.09 & --  /1 & $+$0.00 & 0.06/6 & $+$0.23 & 0.06/15 & $+$0.27 & 0.10/5 & $-$0.06 & 0.07/12 & $-$0.15 & 0.10/ 6 & $-$0.12 & --  /1 \\
16231672-2634279  & $-$0.05 & --  /1 & $+$0.01 & 0.05/6 & $+$0.24 & 0.05/15 & $+$0.27 & 0.07/5 & $-$0.03 & 0.08/13 & $-$0.14 & 0.14/ 6 & $-$0.13 & --  /1 \\
16232114-2631598  & $+$0.16 & 0.07/6 & $+$0.13 & 0.07/6 & $+$0.32 & 0.08/14 & $+$0.30 & 0.02/4 & $-$0.01 & 0.07/ 9 & $-$0.05 & 0.06/ 7 & $+$0.03 & 0.05/2 \\
16232988-2631490  & $-$0.12 & --  /1 & $-$0.02 & 0.06/6 & $+$0.21 & 0.05/16 & $+$0.28 & 0.10/5 & $-$0.08 & 0.07/13 & $-$0.15 & 0.08/ 6 & $-$0.12 & --  /1 \\
16233020-2633241  & $+$0.06 & 0.08/5 & $+$0.02 & 0.08/5 & $+$0.23 & 0.05/12 & $+$0.29 & 0.16/5 & $-$0.10 & 0.22/12 & $-$0.11 & 0.17/ 6 & $-$0.16 & --  /1 \\
16233067-2629390  & $-$0.10 & 0.14/6 & $-$0.13 & 0.14/6 & $+$0.25 & 0.06/14 & $+$0.27 & 0.17/5 & $-$0.10 & 0.21/10 & $-$0.11 & 0.10/ 6 & $-$0.11 & --  /1 \\
16233667-2630397  & $-$0.04 & --  /1 & $-$0.01 & 0.07/6 & $+$0.22 & 0.04/15 & $+$0.29 & 0.13/5 & $-$0.04 & 0.10/13 & $-$0.14 & 0.11/ 6 & $-$0.07 & --  /1 \\
16233846-2629235  & $-$0.11 & --  /1 & $-$0.02 & 0.03/6 & $+$0.23 & 0.07/15 & $+$0.30 & 0.14/5 & $-$0.07 & 0.07/12 & $-$0.17 & 0.09/ 8 & $-$0.14 & --  /1 \\
\hline
 mean             & $-$0.04 &        & $+$0.01 &        & $+$0.25 &         & $+$0.28 &        & $-$0.05 &         & $-$0.12 &         & $-$0.10\\              
 $\pm$            & 0.02    &        & 0.02    &        & 0.01    &         & 0.01    &        & 0.01    &         & 0.01    &         & 0.01\\              
 rms              & 0.07    &        & 0.07    &        & 0.03    &         & 0.03    &        & 0.05    &         & 0.04    &         & 0.06\\              
\hline
\multicolumn{15}{c}{NGC\,2808} \\

09120665-6450253  & $+$0.03 & 0.05/6 & $+$0.00 & 0.05/6 & $+$0.20 & 0.07/14 & $+$0.24 & 0.11/5 & $-$0.11 & 0.09/13 & $-$0.10 & 0.12/8  & $-$0.12 & 0.02/2  \\
09120251-6451001  & $-$0.03 & --  /1 & $-$0.05 & 0.06/6 & $+$0.18 & 0.07/14 & $+$0.13 & 0.11/5 & $-$0.12 & 0.05/11 & $-$0.11 & 0.13/7  & $-$0.12 & 0.01/2  \\
09120213-6452243  & $+$0.04 & 0.05/6 & $+$0.01 & 0.05/6 & $+$0.14 & 0.06/13 & $+$0.22 & 0.13/5 & $-$0.14 & 0.11/12 & $-$0.16 & 0.09/7  & $-$0.10 & 0.07/2  \\
09114655-6452144  & $-$0.08 & --  /1 & $+$0.01 & 0.09/5 & $+$0.15 & 0.08/15 & $+$0.20 & 0.13/5 & $-$0.07 & 0.09/13 & $-$0.11 & 0.13/6  & $-$0.14 & --  /1  \\
09122027-6448450  & $-$0.15 & --  /1 & $+$0.02 & 0.05/6 & $+$0.17 & 0.05/15 & $+$0.24 & 0.11/5 & $-$0.12 & 0.04/12 & $-$0.12 & 0.10/8  & $-$0.07 & 0.02/2  \\
09123016-6454129  & $-$0.09 & --  /1 & $+$0.06 & 0.03/5 & $+$0.23 & 0.09/16 & $+$0.32 & 0.11/4 & $-$0.07 & 0.07/12 & $-$0.09 & 0.08/7  & $-$0.05 & --  /1  \\
09120852-6449107  & $+$0.12 & 0.06/6 & $+$0.08 & 0.06/6 & $+$0.24 & 0.09/14 & $+$0.34 & 0.09/5 & $-$0.04 & 0.10/13 & $-$0.07 & 0.10/7  & --      &  -- /0  \\
\hline
 mean             & $-$0.02 &        & $+$0.02 &        & $+$0.19 &         & $+$0.24 &        & $-$0.10 &         & $-$0.11 &         & $-$0.10\\              
 $\pm$            & 0.03    &        & 0.02    &        & 0.01    &         & 0.03    &        & 0.01    &         & 0.01    &         & 0.01\\              
 rms              & 0.09    &        & 0.04    &        & 0.04    &         & 0.07    &        & 0.03    &         & 0.03    &         & 0.03\\              
\hline  
\end{tabular}
\end{table}
\end{landscape}

%%%%%%%%%%%%%%%%%%%%%%%%%%%%%%%%%%%%%%%%%%%%%%%%%%%%%%%%%%%%%%%%%%%%

\begin{landscape}
\begin{table}
\center
\caption{Derived chemical abundances for Fe, Co, Ni, Zn, Y, and Ce in M\,4 and NGC\,2808 AGBs.
  \label{tab:abu3}}
\scriptsize
\begin{tabular}{c ccc ccc ccc ccc cc}
\\
\hline
\hline
ID (2MASS)&$\rm {[Fe/H]}$\,{\sc i} & rms/\#  &$\rm {[Fe/H]}$\,{\sc ii} & rms/\#  &$\rm {[Co/Fe]}$ & rms/\# &$\rm {[Ni/Fe]}$ & rms/\# &$\rm {[Zn/Fe]}$ & rm/\# &$\rm {[Y/Fe]}$\,{\sc ii} & rms/\# &$\rm {[Ce/Fe]}$\,{\sc ii} & rms/\#  \\
\hline
\multicolumn{15}{c}{M\,4} \\

16233142-2633110   & $-$1.18 & 0.08/75 & $-$1.14 & 0.09/6 &  $-$0.03 & 0.02/4 & $-$0.01 & 0.08/28 & $+$0.19 & --  /1 & $+$0.26 & 0.16/4 & $+$0.13 & --/1 \\
16233477-2631349   & $-$1.22 & 0.08/68 & $-$1.20 & 0.08/8 &  $+$0.07 & --  /1 & $+$0.01 & 0.10/28 & $+$0.09 & --  /1 & $+$0.39 & 0.09/4 & $+$0.31 & --/1 \\
16233535-2632225   & $-$1.26 & 0.08/88 & $-$1.24 & 0.10/8 &  $-$0.07 & 0.06/4 & $-$0.03 & 0.08/30 & $+$0.16 & --  /1 & $+$0.29 & 0.18/3 & $+$0.17 & --/1 \\
16233614-2632015   & $-$1.12 & 0.07/92 & $-$1.08 & 0.07/9 &  $-$0.05 & 0.06/4 & $-$0.03 & 0.06/32 & $+$0.19 & --  /1 & $+$0.32 & 0.15/4 & $+$0.09 & --/1 \\
16233741-2638238   & $-$1.22 & 0.08/67 & $-$1.19 & 0.18/9 &  $-$0.05 & 0.05/3 & $-$0.03 & 0.10/24 & $+$0.34 & 0.07/2 & $+$0.23 & 0.11/4 & $+$0.01 & --/1 \\
16234085-2631215   & $-$1.18 & 0.09/84 & $-$1.15 & 0.08/8 &  $-$0.07 & 0.09/4 & $-$0.02 & 0.07/31 & $+$0.25 & --  /1 & $+$0.36 & 0.15/3 & $+$0.35 & --/1 \\
16234268-2631209   & $-$1.07 & 0.09/45 & $-$1.04 & 0.14/8 &  $-$0.11 & --  /1 & $-$0.01 & 0.08/14 & $+$0.24 & --  /1 & $+$0.38 & 0.06/4 & $+$0.25 & --/1 \\
16234740-2631463   & $-$1.28 & 0.08/89 & $-$1.25 & 0.10/9 &  $-$0.03 & 0.04/4 & $-$0.02 & 0.06/30 & $+$0.31 & 0.18/2 & $+$0.21 & 0.14/3 & $+$0.15 & --/1 \\
16235375-2634426   & $-$1.21 & 0.08/87 & $-$1.18 & 0.07/9 &  $-$0.03 & 0.06/4 & $-$0.02 & 0.07/28 & $+$0.23 & --  /1 & $+$0.27 & 0.13/4 & $+$0.11 & --/1 \\
16240858-2624552   & $-$1.27 & 0.07/85 & $-$1.24 & 0.08/9 &  $-$0.06 & 0.08/4 & $-$0.04 & 0.06/30 & $+$0.20 & 0.03/2 & $+$0.16 & 0.12/4 & $+$0.07 & --/1 \\
16231672-2634279   & $-$1.24 & 0.08/86 & $-$1.20 & 0.07/9 &  $-$0.07 & 0.05/3 & $-$0.02 & 0.07/30 & $+$0.17 & --  /1 & $+$0.25 & 0.11/4 & $+$0.13 & --/1 \\
16232114-2631598   & $-$1.14 & 0.07/67 & $-$1.11 & 0.07/9 &  $-$0.03 & 0.08/2 & $+$0.00 & 0.09/28 & $+$0.16 & --  /1 & $+$0.33 & 0.13/4 & $+$0.18 & --/1 \\
16232988-2631490   & $-$1.28 & 0.07/83 & $-$1.25 & 0.08/9 &  $-$0.04 & 0.05/4 & $-$0.04 & 0.07/31 & $+$0.33 & 0.15/2 & $+$0.28 & 0.16/4 & $+$0.13 & --/1 \\
16233020-2633241   & $-$1.19 & 0.07/66 & $-$1.15 & 0.10/7 &  $-$0.02 & 0.04/3 & $-$0.01 & 0.07/27 & $+$0.44 & 0.07/2 & $+$0.21 & 0.15/3 & $+$0.08 & --/1 \\
16233067-2629390   & $-$1.18 & 0.07/61 & $-$1.15 & 0.15/9 &  $-$0.02 & 0.05/3 & $+$0.01 & 0.11/23 & $+$0.30 & 0.01/2 & $+$0.14 & 0.09/4 & $+$0.02 & --/1 \\
16233667-2630397   & $-$1.24 & 0.08/82 & $-$1.21 & 0.07/9 &  $-$0.04 & 0.05/4 & $-$0.01 & 0.07/27 & $+$0.23 & --  /1 & $+$0.30 & 0.12/4 & $+$0.06 & --/1 \\
16233846-2629235   & $-$1.17 & 0.08/78 & $-$1.14 & 0.06/8 &  $-$0.04 & 0.02/3 & $-$0.03 & 0.09/30 & $+$0.16 & --  /1 & $+$0.17 & 0.10/4 & $+$0.13 & --/1 \\
\hline                                  
 mean              & $-$1.20 &         & $-$1.17 &        & $-$0.04  &        & $-$0.02 &         & $+$0.23 &        & $+$0.27 &         & $+$0.14\\              
 $\pm$             & 0.01    &         & 0.01    &        & 0.01     &        & 0.01    &         & 0.02    &        & 0.02    &         & 0.02\\              
 rms               & 0.06    &         & 0.06    &        & 0.04     &        & 0.01    &         & 0.08    &        & 0.07    &         & 0.09\\              
\hline
\multicolumn{15}{c}{NGC\,2808} \\       
                                        
09120665-6450253   & $-$1.30 & 0.07/68 & $-$1.27 & 0.06/9 &  $-$0.02 & 0.07/3 & $-$0.07 & 0.07/25 & $-$0.01 & --  /1 & $-$0.13 & 0.15/3 & $-$0.17 & --/1  \\
09120251-6451001   & $-$1.31 & 0.08/71 & $-$1.24 & 0.07/9 &  $-$0.17 & 0.07/2 & $-$0.10 & 0.09/26 & $+$0.13 & 0.22/2 & $-$0.16 & 0.06/4 & $-$0.11 & --/1  \\
09120213-6452243   & $-$1.30 & 0.07/67 & $-$1.27 & 0.05/8 &  $-$0.12 & 0.18/2 & $-$0.07 & 0.13/26 & $+$0.03 & --  /1 & $-$0.04 & 0.12/3 & $-$0.13 & --/1  \\
09114655-6452144   & $-$1.25 & 0.08/81 & $-$1.21 & 0.06/9 &  $-$0.15 & 0.00/3 & $-$0.13 & 0.09/27 & $+$0.08 & --  /1 & $-$0.01 & 0.14/3 & $-$0.15 & --/1  \\
09122027-6448450   & $-$1.32 & 0.09/81 & $-$1.29 & 0.06/9 &  $-$0.17 & 0.01/4 & $-$0.10 & 0.09/25 & $+$0.14 & --  /1 & $-$0.03 & 0.10/4 & $-$0.06 & --/1  \\
09123016-6454129   & $-$1.20 & 0.08/75 & $-$1.17 & 0.09/9 &  $-$0.13 & 0.04/2 & $-$0.12 & 0.08/23 & $+$0.21 & 0.16/2 & $+$0.11 & 0.10/4 & $+$0.08 & --/1  \\
09120852-6449107   & $-$1.19 & 0.08/80 & $-$1.15 & 0.07/9 &  $-$0.06 & 0.01/3 & $-$0.04 & 0.11/19 & $+$0.20 & 0.04/2 & $+$0.07 & 0.16/3 &    --   & --/0  \\
\hline                                  
 mean              & $-$1.27 &         & $-$1.23 &        & $-$0.12  &        & $-$0.09 &         & $+$0.11 &        & $-$0.03 &         & $-$0.09\\              
 $\pm$             & 0.02    &         & 0.02    &        & 0.02     &        & 0.01    &         & 0.03    &        & 0.04    &         & 0.04\\              
 rms               & 0.05    &         & 0.05    &        & 0.05     &        & 0.03    &         & 0.08    &        & 0.09    &         & 0.08\\              
\hline
\end{tabular}
\end{table}
\end{landscape}

\section{Summary and conclusions}\label{summary}
We provide a photometric and spectroscopic investigation of multiple
populations along the AGB of the Galactic GCs NGC\,2808 and M\,4. 
Our study is based on: {\it (i)} high-resolution spectroscopy from
FLAMES@VLT; {\it (ii)} multi-wavelength photometry from the {\it HST} UV
survey of Galactic GCs and from ground-based telescopes; and {\it (iii)}
proper motions derived by combining stellar positions from Gaia DR1
and positions derived from images collected with WFI@MPI2.2m
telescope. 

In NGC\,2808 we have identified three main stellar populations of
AGB-1, AGB-2, and AGB-3 stars that populate three AGB sequences in
the $m_{\rm F438W}$ vs.\,$C_{\rm F275W,F336W,F438W}$ pseudo-CMD and in
the $\Delta_{C \rm F275W, F336W, F438W}$ vs.\,$\Delta_{\rm F275W,
  F814W}$ pseudo two-color diagram, or `chromosome map'. The three
populations of AGB-1, AGB-2, and AGB-3 stars include the 41\%, 32\%,
and 27\% of the total number of AGB stars and  have different O, Na,
Mg, and Al abundances.  
 This evidence of multiple populations of AGB stars in NGC\,2808
  adds to the recent finding by Wang et al.\,(2016) based on the distribution
of Na in the same cluster.
 By combining information from this paper and from the literature we
 followed multiple stellar populations along the different
 evolutionary phases, from the MS to the HB and AGB of NGC\,2808.  

Recent papers show that NGC\,2808 hosts five main populations, namely
A--E, that have been detected along the MS and the RGB by using
multi-wavelength photometry and correspond to stellar populations with
different helium and light-element abundance (Milone et al.\,2015a;
Carretta\,2015). 
Unfortunately, there are no spectroscopic studies on population-A stars. 
On the AGB, we find in this paper that AGB-1 stars mostly correspond
to population B, while AGB-2 stars are the progeny of the RGB-C. 
Population-D stars are enhanced in helium up to $Y \sim 0.32$ and have
low oxygen and high sodium abundance. 
We have shown that population-D stars climb
the AGB and define the sequence of AGB-3 stars in the $m_{\rm F438W}$
vs.\,$C_{\rm F275W,F336W,F438W}$ pseudo-CMD. 

We did not find any spectroscopic evidence for population-E stars with
extreme helium and light-element abundance along the AGB although the
small number of analyzed stars prevents us from strong conclusions on
the basis of spectroscopy only. However, this idea is strengthened
  by the analysis of the 'chromosome map' of AGB stars: by comparing the relative
numbers of stars along the distinct AGBs and RGBs we concluded that
the fraction of AGB-3 stars with respect to the total number of AGB stars
is not consistent with the presence of population-E along the AGB. 
 The possibility that population-E stars avoid the AGB phase is
 further supported by the presence of evolved stars that are clearly
 visible in the $m_{\rm F275W}$ vs.\,$m_{\rm F275W}-m_{\rm F814W}$ CMD
 of Figure~\ref{fig:NGC2808target} (blue diamonds) and that have been interpreted by
 Castellani et al.\,(2003) as ``AGB manque''' stars. 
 Since population-E stars have extreme helium abundance ($Y \sim
 0.38$, D'Antona et al.\,2005; Piotto et al.\,2007; Milone et
 al.\,2015a), our findings support the prediction from stellar
 evolution that He-rich stars in stellar populations avoid the AGB
 phase and evolve as AGB-manque' stars
 (e.g.\,Greggio \& Renzini 1990; D'Cruz et al.\,2000; Brown et
 al.\,2001; Moehler et al.\,2004; Gratton et al.\,2010; Chantereau et al.\,2016). 
  
 Specifically, we note that NGC\,2808 is considered a quite young
   GC (age=11.5$\pm$0.75 Gyr, Dotter et al.\,2010; Milone et
   al.\,2014). The lack of stars with extreme helium abundance along
   the AGB of NGC\,2808 would be in agreement with the conclusion by
   Charbonnel \& Chantereau\,(2016), who predict that the internal AGB
   helium spread of a GC with [Fe/H]=-1.15 and age=11.5 is smaller
   than $\Delta Y \sim$0.09. 
   The GC M\,4 has similar metallicity and age (12.50$\pm$0.50~Gyr) as NGC\,2808
  (although NGC\,2808 seems slightly younger, Mar{\'{\i}}n-Franch et
  al.\,2009).
In contrast, in the context of multiple populations, M\,4 looks much
less complex than NGC\,2808. This cluster hosts two main populations
of stars with 
 different C, N, O, Na abundance that have been observed along the MS
 and RGB (e.g.\,Marino et al.\,2008; Piotto et al.\,2015).  
Na-poor/O-rich stars are slightly enhanced in
 helium by $\Delta Y \sim 0.02$ (Nardiello et al.\,2015) with respect
 to the primordial value  and populate the red HB while the blue HB
 host stars more-depleted in 
 oxygen and with higher in sodium (Marino et al.\,2011). 
Despite the possible presence of systematics between the abundances of
AGB and RGB stars, we find that the
chemical abundances dispersions of AGB stars in M\,4 are  
  not consistent with a simple stellar population and provide both photometric
 and spectroscopic evidence that stars belonging to different populations ascend the
 AGB of this cluster.

 In conclusion, while in NGC\,2808 the extremely He-rich (Na-rich,
  O-poor) population of the RGB very likely misses the AGB
  phase, in M\,4 we do not find any strong evidence for a lack of some
  of the RGB populations on the AGB.
  Except for the lack of the extremely He-enhanced population of
  NGC\,2808, the number ratios of second population AGB stars are similar to
  those observed on the RGB in both GCs. These results suggest that
  only a high level of He enrichment, like in the extreme population of
  NGC\,2808, is able to make a star avoiding the AGB phase. 
  At a given metallicity and age, He seems to be the main parameter
  controlling evolution towards the AGB.    
  
%__________________________________________________________________
 \begin{acknowledgements}
   We are grateful to the referee for several suggestion that have
   improved the quality of this manuscript. We thank P.\,B.\,Stetson
   who has kindly provided us the ground-based photometric catalogs of
   M\,4 and NGC\,2808, Ben MacClean for supplying positions of his
   M4 AGB stars to enable a cross-match with our sample, and Simon
   Campbell for useful discussion. 
This work has been supported by the Australian Research Council
(grants DE160100851, DE150101816, DP150100862, FT140100554, FL110100012).
% AFM and APM and HJ acknowledge support by the Australian Research
% Council through Discovery Early Career Researcher Awards DE160100851
% and DE150101816 and Discovery project DP150100862. MA has been
% supported by a Laureate Fellowship awarded by the Australian
% Research Council (grant FL110100012).  
\end{acknowledgements}
%__________________________________________________________________
\bibliographystyle{aa}

\end{document}